\documentclass[sigconf,10pt]{acmart}
    
\renewcommand\footnotetextcopyrightpermission[1]{}
\author{Shan Jiang}
 \affiliation{
   \institution{University of Science and Technology of China}
   \country{}
 }
\email{jiangshan@mail.ustc.edu.cn}

\author{Zhenhua Han}
 \affiliation{
   \institution{University of Science and Technology of China}
   \country{}
 }
 \email{hzhua201@gmail.com}
 \authornotemark[1]

 \author{Haisheng Tan}
 \affiliation{
   \institution{University of Science and Technology of China}
   \country{}
 }
 \email{hstan@ustc.edu.cn}

 \author{Xinyang Jiang}
 \affiliation{
   \institution{Microsoft Research Asia}
   \country{}
 }
 \email{xinyangjiang@microsoft.com}

 \author{Yifan Yang}
 \affiliation{
   \institution{Microsoft Research Asia}
   \country{}
 }
 \email{yifanyang@microsoft.com}

 \author{Xiaoxi Zhang}
 \affiliation{
   \institution{Sun Yat-sen University}
   \country{}
 }
 \email{zhangxx89@mail.sysu.edu.cn}

 \author{Hongqiu Ni}
 \affiliation{
   \institution{University of Science and Technology of China}
   \country{}
 }
\email{291507758@qq.com}

\author{Yuqing Yang}
 \affiliation{
   \institution{Microsoft Research Asia}
   \country{}
 }
 \email{yuqing.yang@microsoft.com}

 \author{Xiang-Yang Li}
 \affiliation{
   \institution{University of Science and Technology of China}
   \country{}
 }
\email{xiangyangli@ustc.edu.cn}

\acmConference []{}{}{}
\renewcommand{\shortauthors}{}

\newcommand{\awRe}{\texttt{River}}
\newcommand{\randomRe}{\texttt{randomRe}}

\newcommand{\ie}{\emph{i.e.}}
\newcommand{\eg}{\emph{e.g.}}

\authornote{The author had the honor of conducting a visiting research at University of Science and Technology of China.}

\usepackage{microtype}
\usepackage{graphicx}
\usepackage{booktabs} 
\usepackage{caption}
\usepackage{subcaption}

\usepackage[ruled,linesnumbered]{algorithm2e}
\usepackage{siunitx}
\usepackage{multirow}
\usepackage{url}

\usepackage{amsmath}
    
\usepackage{amssymb}

\begin{document}


\title{Real-Time Neural-Enhancement for Online Cloud Gaming }

\begin{abstract}
Online Cloud gaming demands real-time, high-quality video transmission across variable wide-area networks (WANs). Neural-enhanced video transmission algorithms employing super-resolution (SR) for video quality enhancement have effectively challenged WAN environments. However, these SR-based 
methods require intensive fine-tuning for the whole video, making it infeasible in diverse online cloud gaming. 

To address this, we introduce \awRe{}, a cloud gaming delivery framework designed based on the observation that video segment features in cloud gaming are typically repetitive and redundant. This permits a significant opportunity to reuse fine-tuned SR models, reducing the fine-tuning latency of minutes to query latency of milliseconds.
To enable the idea, we design a practical system that addresses several challenges, such as model organization, online model scheduler, and transfer strategy.
\awRe{} first builds a content-aware encoder that fine-tunes SR models for diverse video segments and stores them in a lookup table.  
When delivering cloud gaming video streams online, \awRe{} checks the video features and retrieves the most relevant SR models to enhance the frame quality.  
Meanwhile, if no existing SR model performs well enough for some video segments, \awRe{} will further fine-tune new models and update the lookup table. 
Finally, to avoid the overhead of streaming model weight to the clients, \awRe{} designs a prefetching strategy that predicts the models with the highest possibility of being retrieved. 

Our evaluation based on real video game streaming demonstrates \awRe{} can reduce redundant training overhead by 44\% and improve the Peak-Signal-to-Noise-Ratio by 1.81dB compared to the SOTA solutions. Practical deployment shows \awRe~meets real-time requirements, achieving approximately 720p 20fps on mobile devices.
\end{abstract}

\maketitle

\section{Introduction}

High-quality gaming requires real-time rendering, 
which is impractical without power-intensive and high-end hardware, making online cloud gaming a preferable choice, especially for mobile users. Online Cloud gaming \cite{huang2013cloudgaming,shea2013cloudgaming} can be considered as a special video transmission task. It typically consists of a server and a client, where the server is responsible for rendering the game visuals interactively and transmitting them back to the client in real-time through peer-to-peer communication frameworks like \emph{WebRTC}~\cite{WebRTC}.  
In fluctuating network conditions as wide-area networks (WANs), \emph{WebRTC} frequently uses adaptive dynamic bitrate encoding to manage the dynamic transmission of video streams \cite{dobrian2011understanding,pires2014dash,meng2023enabling,song2023halp}, leading that the clients might receive video streams of lesser resolution and bitrate.  
This is the main reason why cloud gaming, such as OnLive \cite{OnLive}, can not be widely promoted, namely, due to high dependence on network conditions.

\begin{figure}[htbp] 
    \centering 
    \includegraphics[width=0.8\linewidth]{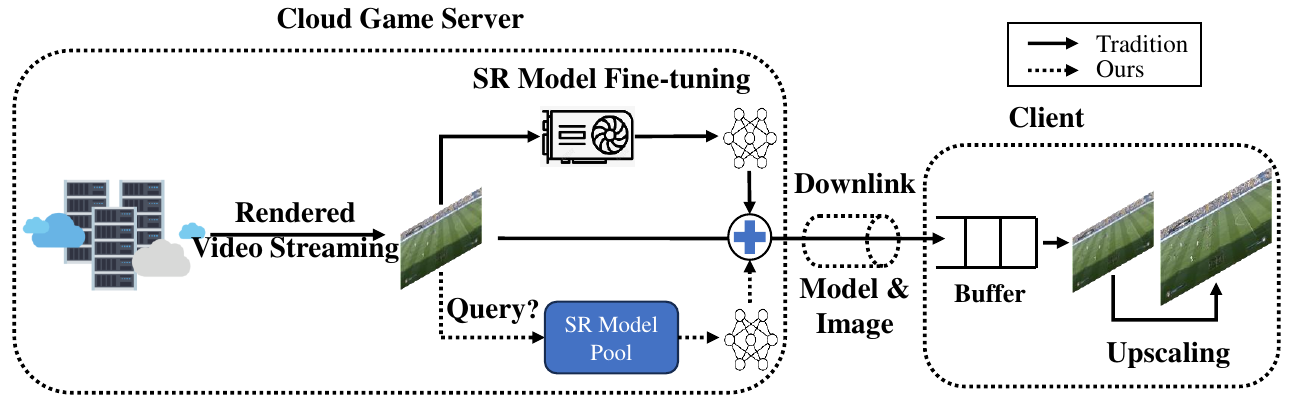} 
    \caption{Neural-enhanced Video Delivery Frameworks.}
    \Description{Neural-enhanced Video Delivery Frameworks.}
    \label{fig.intro} 
\end{figure}

In order to find a balance between computation and networking of video delivery, a list of neural-enhanced video delivery frameworks~\cite{kim2020LiveNAS,yeo2018NAS,liu2021CaFM,hu2019dejavu,baek2021dcsr} are proposed. As shown in the solid part of \autoref{fig.intro}, they typically consist
of three main components. On the server side, independent super-resolution (SR) models are fine-tuned for each video segment online or offline, called content-aware fine-tuning. Then, the low-quality video segments are transmitted to the client along with the fine-tuned SR models over the network. Finally, on the client side, the SR models are executed to enhance the low-resolution images into high-resolution images to improve the quality of service (QoS). This solution is more friendly to low-bandwidth users who have additional computing power.

Unfortunately, fine-tuning models for each video segment is infeasible for online cloud gaming. First, online cloud gaming is an interactive service requiring a latency of tens of milliseconds. Compared to the prior-unknown streaming in video-on-demand (VoD), the game video frames for streaming are generated via real-time rendering. Existing approaches~\cite{yeo2018NAS,kim2020LiveNAS,li2022emt,chen2023vichaser} that endeavor to fine-tune an SR model during the streaming video is not feasible for online gaming. Besides, adding a model training step into the pipeline introduces a significant training delay (usually in minutes).
Second, the video frames of cloud gaming change very rapidly over time, such as transitioning from indoor to outdoor scenes, which will undoubtedly further reduce the benefits of fine-tuning. Finally, in contrast to one-to-many video streaming (\eg, broadcasting), cloud gaming's video streaming is one-to-one, which requires much more GPU resources for fine-tuning SR models to handle the many concurrent streaming sessions. Therefore, it poses an important question: \emph{is it possible for online cloud gaming to benefit from the video quality enhanced by SR models while satisfying the requirement of user-perceived latency and deployment cost?}

In this paper, we give an affirmative answer to the above question. Through analyzing the fine-tuning process for cloud gaming, we identify a key observation that training independent models based on the whole data
is repetitive and redundant. This is mainly because of the similarity among video streams received by different clients and the temporal similarity of video streams received by the same client. Due to such similarity, repetitive training of these similar video segments is often unnecessary. In response, we propose querying whether existing well-trained SR models can meet the image enhancement requirements of the current video stream, as shown in \autoref{fig.intro}, prospective to significantly reduce redundant training costs and ensure a certain level of image enhancement performance. However, this solution will introduce a large pool of SR models to handle the various clusters of video features. We can't download all models to clients, as the video streaming latency will inevitably increase if too much network bandwidth is occupied to transmit the model weights. It leads to a trade-off between video streaming and model streaming. 
In summary, we aspire to devise a cloud gaming transmission system that enhances video quality through content-aware retrieval to meet the stringent real-time requirements of cloud gaming and address the issues of computational resource wastage inherent in existing systems. Specifically, we try to address the following key challenges:

\begin{itemize}
\item How to build the model pool that can efficiently handle diverse video content when retrieving SR models?
\item How to determine the optimal reusable model from the model pool for the current video frames and handle unseen video streaming?
\item Due to limited 
bandwidth, how to balance the transmission of video streaming and the model delivery?
\end{itemize}




In this work, we propose a \underline{R}etr\underline{i}eval-based \underline{v}ideo \underline{e}nhancement f\underline{r}amework for cloud gaming, \awRe, to reduce training costs and latency as well as enhance video quality at clients.
Our main contributions can be summarized as follows.

\begin{itemize}
\item First, we emphasize that fine-tuning each segment of a cloud gaming video is unnecessary, specifically when they reflect repetitive patterns in time and space. Our experiments demonstrate that reusing models can achieve a decent Quality of Service (Section \ref{sec:motivation}).



\item Second, we propose a cloud gaming delivery framework named \awRe~with a content-aware model encoder for efficient model reuse, an online scheduler for optimal model selection and model fine-tuning based on the video stream, and a model prefetching strategy designed to minimize bandwidth impact during cloud game video streaming (Section \ref{sec:system}).

\item Third,  we implemented \awRe~and extensive experiments demonstrate that our framework reduces the redundant training overhead by 44\% and improves the Peak Signal-to-Noise Ratio (PSNR) by 1.81dB on average compared with the SOTA methods. Further,  \awRe~meets real-time requirements when practically deploying it on mobile devices (Section \ref{sec:impl}, \ref{sec:evaluation}).
\end{itemize}

\section{Background}
\subsection{Super-resolution}
\noindent\textbf{Super-resolution} (SR) is a field of study that has seen significant advancements with the advent of Deep Learning \cite{dong2015srcnn,kim2016vdsr,lim2017edsr,yu2020wdsr,liang2021swinir,chen2023hat}. It focuses on enhancing the low-resolution images (LR) by the upscaling factor to the high-resolution images (HR), called image enhancement.

\noindent\textbf{Peak Signal-to-Noise Ratio} (PSNR) is a commonly used metric in image processing, 
which is calculated based on the pixel-wise difference between the original and reconstructed images (\autoref{eq.psnr}) and expressed in decibels.
\begin{equation}
    PSNR = 10 \cdot \log_{10}\left(\frac{{MAX}^2}{{MSE}}\right)
    \label{eq.psnr}
\end{equation}
where $MAX$ is the maximum possible pixel value (\eg, 255 for an 8-bit image), and $MSE$ is the Mean Squared Error, calculated as the average of the squared pixel-wise differences between the reference and enhanced images. The higher the PSNR, the better the quality of the enhanced image.

\subsection{Neural-enhanced Video Delivery}

Neural-enhanced video transmission algorithms are proposed to address fluctuating video stream quality due to varying super-resolution networks. NAS\cite{yeo2018NAS} is the first framework that considers using DNN models to overfit each video block to ensure reliability and performance in video delivery. Dejavu\cite{hu2019dejavu} proposes encoding repeated video meetings based on the sender's historical video conference sessions and sharing them in advance with the recipient. LiveNAS\cite{kim2020LiveNAS} employs online learning to maximize the quality gain and dynamically adjusts the resource use to the quality improvement. CaFM\cite{liu2021CaFM} studies the relation between models of different chunks and designs a handcrafted layer to compress these models for neural video delivery. 

However, these systems are primarily designed for live streaming or video-on-demand (VoD), where multiple users can watch the same video source. An overfitted model for each video segment can still benefit many receivers. However, cloud gaming adopts a peer-to-peer video streaming approach, where there is only one user receiving from one video source, rendering existing approaches too costly. Next, we discuss the challenges and opportunities of neural-enhanced video streaming for cloud gaming.

\section{Motivation}\label{sec:motivation}
In this section, we highlight the shortcomings of current neural-enhanced video streaming solutions, i.e., the heavy fine-tuning overhead and the degraded image enhancement effect due to visual fluctuations specifically when facing the \textit{diverse} and \textit{online} cloud gaming scenarios. We investigate the temporal and spatial redundancy in video streaming, and point out that it is not necessary to fine-tune models for each video segment in cloud gaming.

\subsection{Heavy Training Cost and Latency}
SR models need to be fine-tuned for video delivery to ensure QoS. \cite{kim2020LiveNAS} claims that most benefits of online training
come from the first few cycles (about $40$ training epochs). Despite this, the training overhead for video delivery is still excessively high, especially in cloud gaming, where the server has to render and generate frames for players.
Moreover, unlike live streaming or video on demand, cloud gaming is one-to-one, requiring handling a larger number of video streams, each of which needs an independent model trained. Besides, the content-aware models are always single-use in the existing framework, such as NAS\cite{yeo2018NAS} and CaFM\cite{liu2021CaFM}, resulting in a substantial waste of GPU utilization.

\begin{table}[htbp]
\caption{Training delay for fine-tuning on a 10-second video segment with an A100 GPU.}
\label{tbl.training_cost}
\begin{tabular}{cccccc}
\toprule
Model & Scale & Size & FLOPs & Training Cost \\
\midrule
\multirow{2}{*}{NAS\cite{yeo2018NAS}} & x2 & 0.5MB & 0.54G & 4min 18s \\
~ & x4 & 2.1MB & 0.81G & 4min 57s \\
\hline
\multirow{2}{*}{WDSR\cite{yu2020wdsr}} & x2 & 4.6MB &4.85G & 13min 8s \\
~ & x4 & 4.7MB & 1.23G & 5min 46s \\
\hline
\multirow{2}{*}{EDSR\cite{lim2017edsr}} & x2 & 5.3MB &5.62G & 13min 27s \\  
~ & x4 & 5.9MB & 2.03G & 6min 33s \\
\bottomrule
\end{tabular}
\end{table}

\autoref{tbl.training_cost} shows the training delay required for fine-tuning different SR models on an A100(80GB) GPU for a 10-second video segment during $40$ training epochs. We divide a frame into patches, using a patch size of $64\times64$ for the scaling factor of $2$ and $32\times32$ for $4$ to accommodate HR video size. Meanwhile, there are other loads on the GPU as well, such as rendering game frames. For a 10-second video segment, we spend around $300$ seconds on training to fit the data, nearly 30 times the length of a video stream. 

\subsection{Enhancement Performance Degradation}
\label{sec.motiv.enhancement_performance}
 
Online cloud gaming experiences more frequent variation in scenes and poses greater challenges compared to video live streaming with relatively static scenes or pre-available video-on-demand streams. Meanwhile, online fine-tuning of the SR models will introduce inherent latency, leading to mismatching of the scenes during the training and inference procedure. The more frequently the scenes change, the worse the mismatching, which sometimes causes the catastrophic decline of video enhancement effects.

To demonstrate this, we downloaded a game video of CSGO from YouTube of nearly 20 minutes and fine-tuned its first 100 seconds. 
In the vein of LiveNAS~\cite{kim2020LiveNAS}, after approximately 40 rounds of fine-tuning using an A100 (80GB) GPU (roughly in 600s), the model was validated on the remaining video stream.
In comparison, we employed a generic super-resolution model pre-trained on a standard dataset\cite{DIV2K}. 

\begin{figure}[htbp] 
    \centering 
    \includegraphics[width=0.7\linewidth]{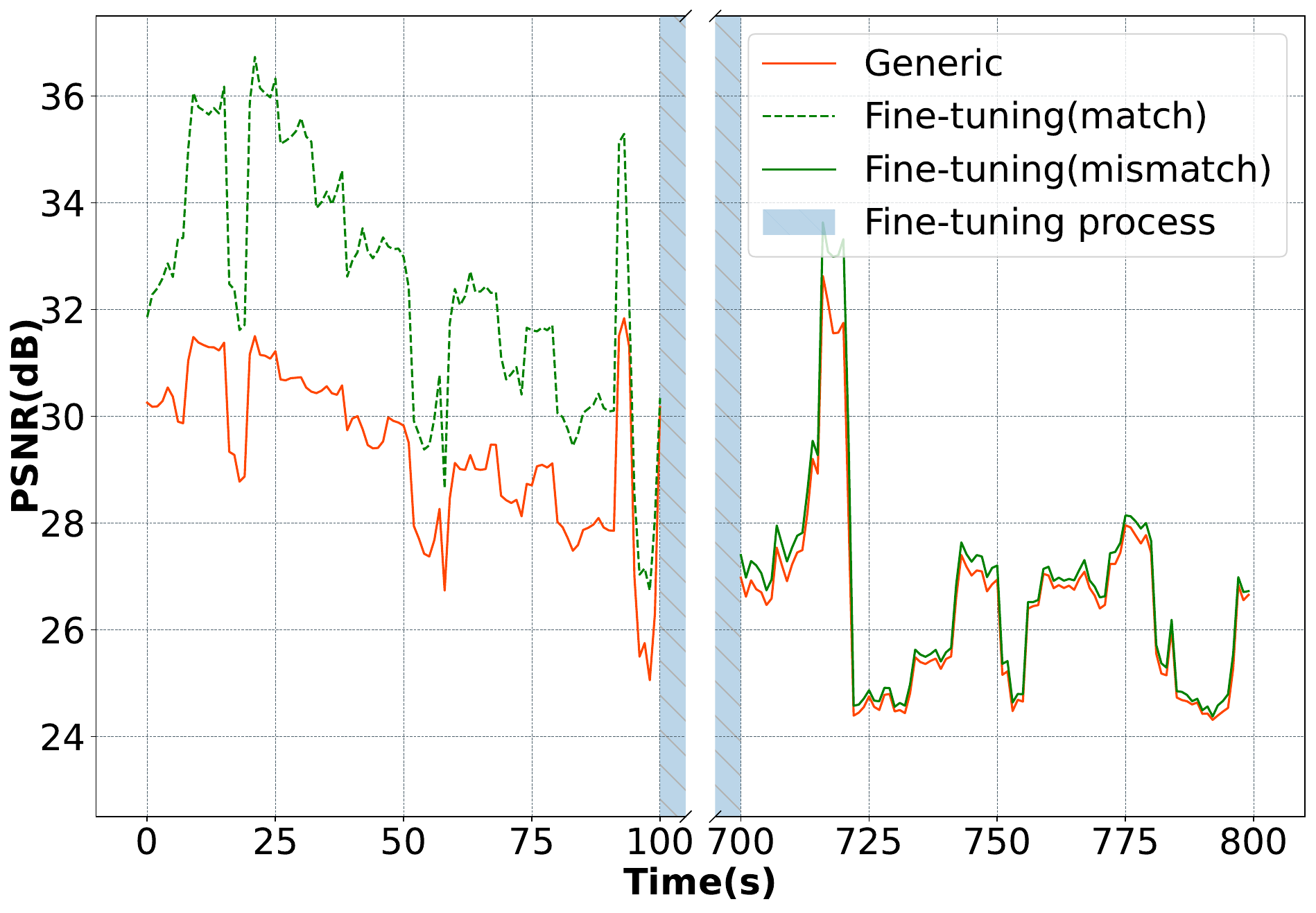} 
    \caption{Performance degradation due to training delay.} 
    \Description{Performance degradation due to training delay.}
    \label{fig.delay_hit_psnr} 
\end{figure}

As shown from 0s to 100s in \autoref{fig.delay_hit_psnr}, fine-tuning the video stream (dotted green line) does yield a satisfactory quality improvement compared to the generic model. However, in practice, the online fine-tuning process does not finish until the 700s. Due to the mismatching in fitting content, the practical performance, represented in the solid green line, is degraded nearly close to the generic.

\subsection{Similarity of Scenes}
\label{sec.motiv.similarity}

\begin{figure*}[t]
    \centering 
    \includegraphics[width=0.7\linewidth]{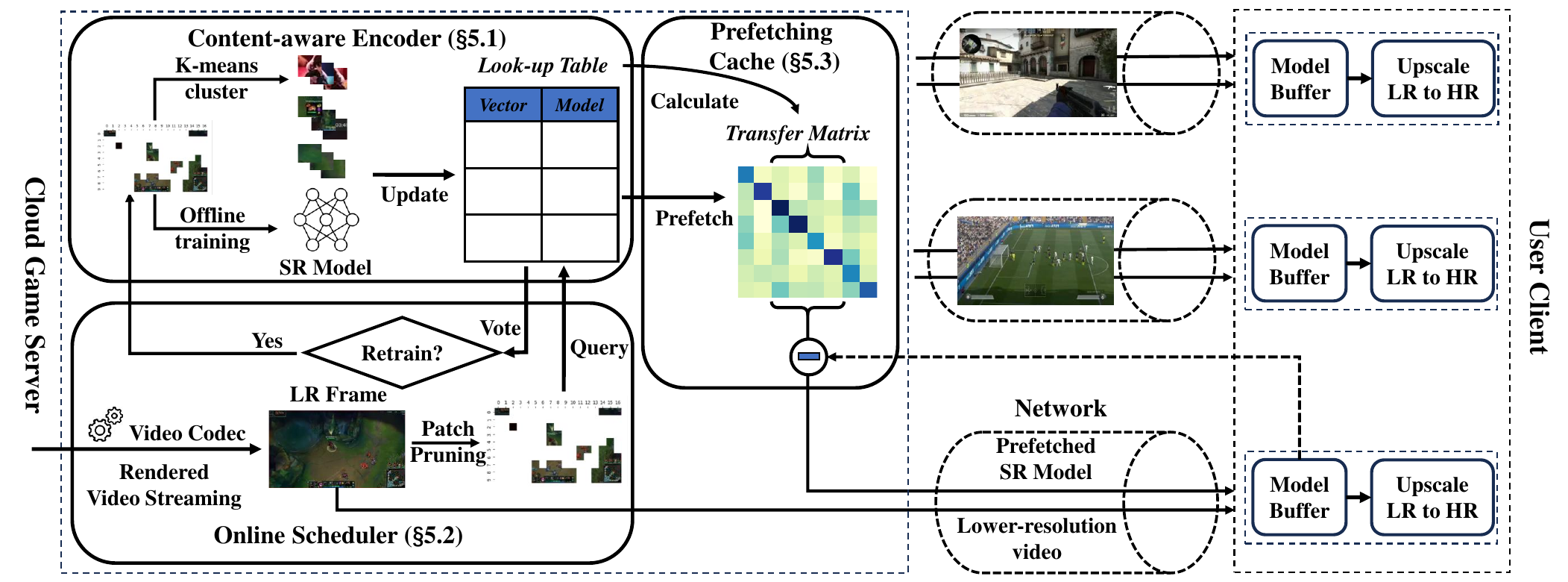} 
    \caption{\awRe~System Overview.} 
    \Description{\awRe~System Overview.} 
    \label{fig.system_overview} 
\end{figure*}

The benefits of content-aware fine-tuning on video streaming mainly come from the model's fitting of content in the videos. For cloud gaming, there is a similarity between video segments transmitted to different users and to the same user at different times.  The SR models fine-tuned on similar segments tend to exhibit comparable performance. It challenges the necessity to train the independent SR models for each video segment in cloud gaming.

\begin{figure}[htbp] 
    \centering 
    \includegraphics[width=0.65\linewidth]{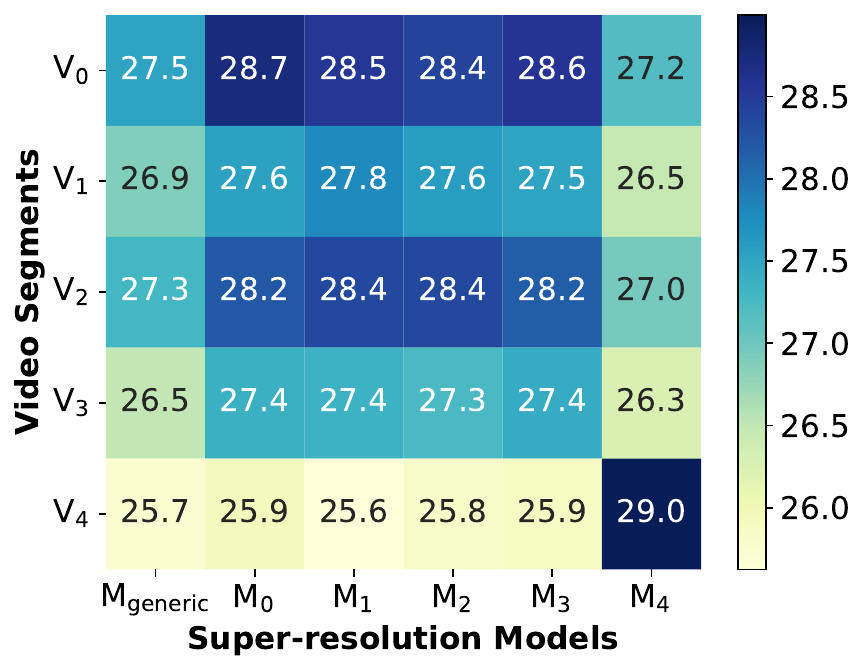} 
    \caption{Comparison of PNSR among different SR models on different video segments.} 
    \Description{Comparison of PNSR among different SR models on different video segments.} 
    \label{fig.pre_exp_with_generic} 
\end{figure}

To prove it, we sampled five 1-minute video segments $\{V_{i}\}_0^4$ from cloud gaming video streams to demonstrate the repetitiveness and redundancy in fine-tuning. $\{V_{i}\}_0^3$ represents video segments from four players in a round of CSGO, while $V_{4}$ is from LoL. We fine-tuned independent SR models $\{M_{i}\}_0^4$ for each segment and validated the image enhancement performance of each model on all video segments. We use a generic SR model, $M_{generic}$, pre-trained on a standard dataset\cite{DIV2K} as the baseline. The results under varying cases are depicted in \autoref{fig.pre_exp_with_generic}. 

The y-axis represents the video segments $\{V_{i}\}_0^4$, and the x-axis represents the generic model and the models $\{M_{i}\}_0^4$ fitted to the video segments. Each cell represents the performance (PSNR) of models on video segments. The darker the color, the better the performance. We can see that the model fine-tuned specifically for the video segment performs the best among the models (\eg, $M_{0}$ on $V_{0}$). Meanwhile, the performance of another model fine-tuned for the same scene (\eg, $\{M_{i}\}_1^3$ on $V_{0}$) generally degrades within 0.1dB to 0.2dB. In other words, there may not be a need to fine-tune models separately for $\{V_{i}\}_0^3$, and instead, selecting one representative existing model may be sufficient. However, if the model and video do not match, a significant decrease in image enhancement will occur (\eg, the difference between the performance of $M_{4}$ on $V_{4}$ and $\{M_{i}\}_0^3$ on $V_{4}$).

Reducing the cost of model fine-tuning for online video streams is a crucial issue for improving the feasibility of the algorithm. For online cloud gaming, even fine-tuning only key frames (such as I frame) has some drawbacks. Because P frames maybe have large differences from I frames in diverse game video streaming, video quality will degrade unexpectedly. The implication of the identified similarity in video segments presents an interesting opportunity for resource optimization - the reuse of previously trained models.
Instead of training a new model for the video segment that has just been rendered, we could identify similar segments and apply a previously fine-tuned model suitable for these segments. This could potentially reduce the need for resource-intensive training processes, leading to significant savings in computational resources and time. Furthermore, this approach can also reduce the latency introduced by the fine-tuning process, as the system can immediately apply a pre-trained model instead of waiting for a new model to be trained. Given the high real-time requirements of cloud gaming, this reduction in latency could greatly improve the experience of online cloud gaming. 

In this paper, we propose \awRe~to exploit the opportunity of reusing models for similar video segments. It requires a robust mechanism for comparing the diverse scene of online video segments and a comprehensive library of pre-trained models that can cater to various types of video content. The detected similarity in video segments opens up new possibilities for improving the efficiency and performance of neural-enhanced video transmission systems.


\section{System Design}\label{sec:system}
To address the shortcomings of existing frameworks in training costs and performance, we propose a novel cloud gaming transmission framework via content-aware retrieval, \awRe, aiming to improve video quality while achieving minimal training overhead and end-to-end latency. The system design overview of \awRe~is illustrated in \autoref{fig.system_overview}.

We design three key modules: \emph{content-aware encoder, online scheduler}, and \emph{prefetching strategy}. The content-aware encoder of \awRe~maintains a set of SR models, each having a vector that represents their preferred video contents. These contents are expected to yield better video quality enhancement with this model (\S\ref{sec.sys.content-aware_encoder}). During the transmission of the compressed cloud game video streaming, the online scheduler first queries a model lookup table to select the most appropriate SR model. This model will be transmitted alongside the video. The scheduler also checks whether the selected model achieves the expected video quality enhancement (\S\ref{sec.sys.scheduler}). If it does not, the video stream will trigger offline training to enrich the SR model pool and improve its content coverage. The lookup table of the content-aware encoder will be updated after the training to include the new SR model for future video delivery. To reduce the bandwidth consumption associated with transmitting model weights, we also design a prefetching strategy. This strategy prefetches and caches the SR model with the highest probability of use (\S\ref{sec.sys.prefetching_cache}). 

\subsection{Content-aware Encoder}
\label{sec.sys.content-aware_encoder}

For efficient and accurate model retrieval, it's crucial to first identify the types of videos that an SR model excels at. As depicted in \autoref{fig.delay_hit_psnr} and \autoref{fig.pre_exp_with_generic}, diversity can exist among different games and even within the same game. The key challenge in reusing SR models lies in organising and representing these models that are suitable for different videos.

To address these challenges, we suggest the creation of a model lookup table that includes models and their encoding. Algorithm \ref{alg.model_encoder} outlines the procedure for building the lookup table. The main idea is to allow the SR model to capture redundancies in the video segment: we can create an appropriate feature extraction algorithm to encode the model based on the segment.

First, we decode the video segment $V_{i}$ into frames (\ie, line \ref{alg.model_encoder.2}). For each frame, we divide it into patches $\{P_{n}\}_0^{N-1}$ of a fixed size (\ie, line \ref{alg.model_encoder.3}-\ref{alg.model_encoder.4}). We use a patch encoder to compute features $embedding_{n}$ for patches $P_{n}$ and $e_{n}$ represents the mean grayscale image obtained after edge detection from the input RGB frame patch, referred to as the edge score. To reduce computational overhead, we focus only on the patches whose edge scores $e_{n}$ exceed a threshold $\lambda$ and include them in the dataset (\ie, line \ref{alg.model_encoder.5}-\ref{alg.model_encoder.10}). We fine-tune the SR model $M_{i}$ on the dataset (\ie, line \ref{alg.model_encoder.12}). For the obtained set of feature vectors $\{embedding_{n}\}_0^M (M < N)$, we use k-means with cosine similarity for clustering to compress redundancy (\ie, line \ref{alg.model_encoder.13}) so that similar frames will have a similar representation. We use the cluster centres $\{\mu_{i}^{0},\dots,\mu_{i}^{K-1}\}$ as the representation of the SR model $M_{i}$ in the lookup table (\ie, line \ref{alg.model_encoder.14}). In summary, the composition of the lookup table is as follows:
\begin{equation}
\centering
    T_{i} \leftarrow < \{\mu_{i}^{0},\dots,\mu_{i}^{K-1}\}, M_{i}> 
\label{eq.model_look-up_table}
\end{equation}
where $T_{i}$ is the item of the lookup table. The model trained on the video segment $V_{i}$ is denoted as $M_{i}$, and $\{\mu_{i}^{0},\dots,\mu_{i}^{K-1}\}$ represents the encoding of $M_{i}$.

\begin{figure}[htbp]
    \begin{minipage}[b]{0.45\linewidth}  
        \vspace{0pt}
        \centering
        \includegraphics[width=\linewidth]{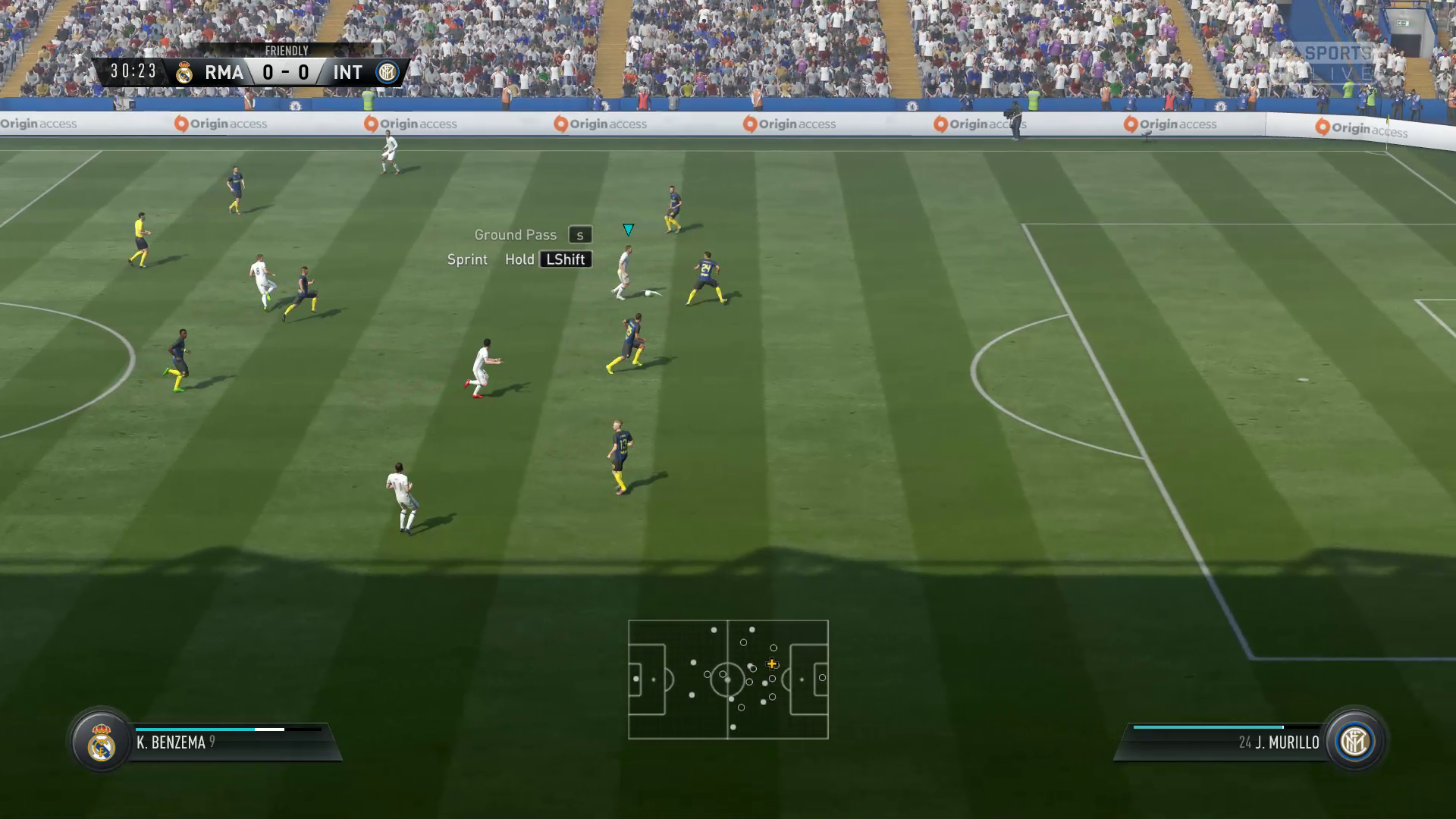}
        \vspace{1pt}
        \subcaption{Original frame.}
        \label{fig.cluster_patch.1}
    \end{minipage} 
    \begin{minipage}[b]{0.45\linewidth}
        \vspace{0pt}
        \centering
        \includegraphics[width=\linewidth]{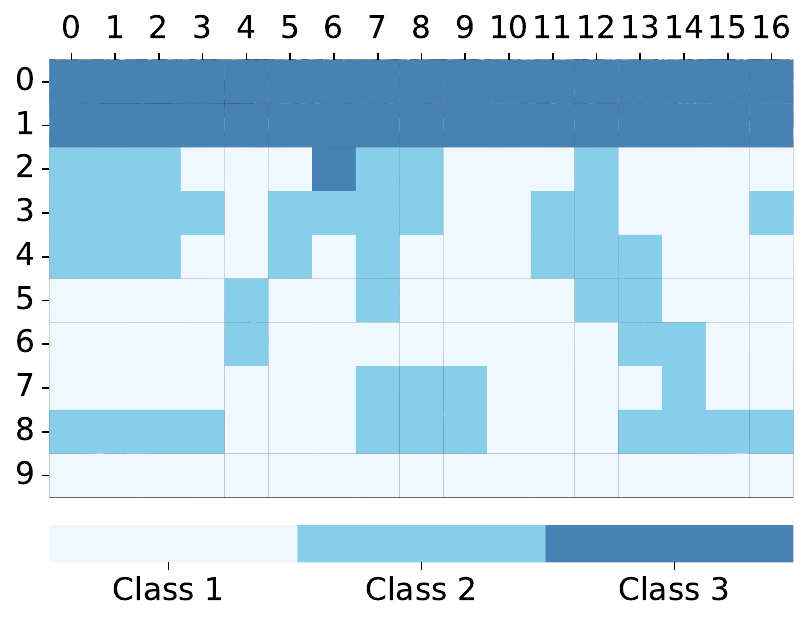}
        \subcaption{Classification mask.}
        \label{fig.cluster_patch.2}
    \end{minipage}
    \caption{(a) The original frame of FIFA17. (b) Different regions are divided into different groups according to our approach.}
    \label{fig.cluster_patch}
\end{figure}

\begin{algorithm}[htb]
    \caption{Update the lookup table}
    \label{alg.model_encoder}
    \KwIn{video segment $V_{i}$, number of cluster centers $K$, edge threshold $\lambda$}
    \KwOut{new item in lookup table $T_{i}$}
    Decode $V_{i}$ into frames\;
    \label{alg.model_encoder.2}
    \For{\textbf{each} frame}{
    \label{alg.model_encoder.3}
        Divide the frame into patches $\{P_{n}\}_0^{N-1}$\;
        \label{alg.model_encoder.4}
        \For{\textbf{each} patch}{
        \label{alg.model_encoder.5}
            \If{$e_{n} > \lambda$}{
                Add $P_{n}$ into the dataset\;
                $embedding_{n} \leftarrow $ Patch Encoder ($P_{n}$)\;
            }
        }
        \label{alg.model_encoder.10}
    }
    Train the model $M_{i}$ on the dataset.\;
    \label{alg.model_encoder.12}
    $\{\mu_{i}^{0},\dots,\mu_{i}^{K-1}\} \leftarrow $ K-means centers of $\{embedding_{n}\}_0^M$ with cosine similarity.\;
    \label{alg.model_encoder.13}
    \Return $T_{i} \leftarrow <\{\mu_{i}^{0},\dots,\mu_{i}^{K-1}\},M_{i}>$.\;
    \label{alg.model_encoder.14}
\end{algorithm}

\autoref{fig.cluster_patch} presents the results of categorizing a single frame from FIFA17 using this approach. We divide the image into patches of equal size (for instance, a 1080p image is divided into 128x128 patches) and use ResNet18 to compute patch embeddings. Subsequently, we apply k-means clustering with three cluster centers. \autoref{fig.cluster_patch.1} displays the original picture. \autoref{fig.cluster_patch.2} illustrates the results post-classification, with different colors indicating different categories. We observe that the background (such as the football field) is grouped into Class 1, the audience on the stands is categorized into Class 3, and the remaining parts (like football players) are assigned to Class 2. We find that this content-aware encoding approach can be applied to a series of videos with similar content, thus a good fit for content-aware SR model retrival.

\subsection{Online Scheduler}
\label{sec.sys.scheduler}
The selection of the optimal SR model from the model lookup table directly impacts the user's Quality of Service (QoS) for online video streams, as mentioned in \autoref{fig.pre_exp_with_generic}. Due to the diversity of cloud gaming, it is essential to determine the best SR model for each frame. Selecting a mismatched model can result in severe performance degradation. Simultaneously, to satisfy real-time requirements\cite{shi2011mobile,meilander2014mobile}, the execution delay of the scheduling algorithm should be minimized.

The online scheduling procedure of \awRe~is depicted as Algorithm \ref{alg.online_schedluer}. The aim is to retrieve the model that was fine-tuned on data similar to the current scene. The pre-trained classification model used in Algorithm \ref{alg.model_encoder} generates similar feature vectors for similar images~\cite{wan2014deep,wang2014learning,van2020scan}. Cosine similarity is used to measure the distance between feature vectors. The current frame, denoted as $F_{i}$, is divided into multiple patches, $\mathcal{D} = \{P_n\}_0^{N-1}$, where $N$ is the total number of patches in the frame (line \ref{alg.online_schedluer.3}). The optimal model for each patch is queried from the lookup table according to \autoref{eq.quary}.
\begin{equation}
\centering
\begin{cases}
    Q(P_{n}) = \arg\max_{j}S_c(embedding_n,\mu_j^k) \\
    j \in (0,\dots,R-1), k \in (0,\dots,K-1)
    \label{eq.quary}
\end{cases}
\end{equation}
Here, the cosine similarity is denoted as $S_{c}$, the feature of $n$-th patch is represented as $embedding_n$, and $\mu_j^k$ is the $k$-th dimension of the $M_j$'s encoding. $K$ represents the number of cluster centers, and $R$ is the number of SR models in the lookup table. In the worst case, each patch may hit a different model. The plurality voting algorithm is used to select the SR model chosen by the most patches and force the entire frame to use the same SR model to prevent excessive model switching costs at the client-side (lines \ref{alg.online_schedluer.7}-\ref{alg.online_schedluer.8}). 

In practice, the number of patches, $N$, might be too large, increasing computational overhead (\eg, a 1080p frame can be split into 170 patches of size 128x128). However, not all patches need to be involved in model selection. Generally, simpler patches benefit less from targeted training than complex ones, as demonstrated in the experiments of Section \ref{sec.patch_pruning}. To reduce the computational cost of the algorithm, we use the score of edge detection\cite{chen2022arm} to prune the patches (lines \ref{alg.online_schedluer.4}-\ref{alg.online_schedluer.6}). The remaining patches are as follows.
\begin{equation}
    \hat{\mathcal{D}} = \{P_{i}|e_{i} > \lambda,P_{i} \in \mathcal{D}\}
    \label{eq.edge_filter}
\end{equation}
Here, $e_{i}$ is the edge score of the i-th patch and $\lambda$ is the threshold. The model to be retrieved for the current frame $F_{i}$ is
\begin{equation}
    M_{F_{i}} = \{M_{j} |j = V_p(Q(P_{n})), P_{n} \in \hat{\mathcal{D}} \} 
    \label{eq.frame_voting}
\end{equation}
where $V_{p}$ is the plurality voting algorithm (lines \ref{alg.online_schedluer.9}-\ref{alg.online_schedluer.11},\ref{alg.online_schedluer.18}). However, the retrieved SR model may not meet the image enhancement requirements, such as low similarity between videos, especially in unseen scenes. If the final vote count does not exceed a threshold (\ie, $\alpha  * count_{p}$), a new content-aware model is trained for the current frame, and the lookup table is updated by Algorithm~\ref{alg.model_encoder} (lines \ref{alg.online_schedluer.14}-\ref{alg.online_schedluer.17}). In practice, we usually fine-tune the model by the unit of video segment. That is when the number of video frames to be fine-tuned in the video segment exceeds  $\alpha$ times the total number of frames, the video segment is fine-tuned.

\begin{algorithm}[htb]
    \caption{Online scheduler}
    \label{alg.online_schedluer}
    \KwIn{video frame $F_{i}$, lookup table $\{T_{i}\}_0^{R-1}$, edge threshold $\lambda$, similarity threshold $\beta$, voting threshold $\alpha$}
    \KwOut{the optimal model retrieved $M_{proxy}$}
    Initialize $count_{p}$ , $vote$\;
    \label{alg.online_schedluer.2}
    Divide frame $F_{i}$ into patches $\{P_{n}\}_0^{N-1}$\;
    \label{alg.online_schedluer.3}
    \For{\textbf{each} patch}{
        \label{alg.online_schedluer.4}
        \If{$e_{n} > \lambda$}{
            $count_{p} \leftarrow count_{p} + 1$\;
            \label{alg.online_schedluer.6}
            $embedding_{n} \leftarrow $  Patch Encoder ($P_{n}$)\;
            \label{alg.online_schedluer.7}
            $ i,s \leftarrow$ index and similarity of the model in $\{T_{i}\}_0^{R-1}$ whose encoding is closest to $embedding_{n}$\;
            \label{alg.online_schedluer.8}
            \If{$s > \beta$}{
                \label{alg.online_schedluer.9}
                $vote[i] \leftarrow vote[i] + 1$\;
            }
            \label{alg.online_schedluer.11}
        }
    }
    \label{alg.online_schedluer.13}
    \If{$\max(vote)$ $ < \alpha * count_{p}$}{
        \label{alg.online_schedluer.14}
        Fine-tune the content-aware model\;
        Update the lookup table\;
    }
    \label{alg.online_schedluer.17}
    \Return $M_{proxy} \leftarrow M_{\arg\max(vote)}$\;
    \label{alg.online_schedluer.18}
\end{algorithm}

\subsection{Prefetching Strategy }
\label{sec.sys.prefetching_cache}
The model queried for each frame in Algorithm~\ref{alg.online_schedluer} represents the best case. However, downloading SR models vies for bandwidth with the video stream, which can potentially increase video streaming latency. The constraint of download bandwidth precludes us from downloading every model. While it may be feasible to miss some model requests, it leads to a decline in video quality. Therefore, it is imperative to devise a rational model delivery algorithm for optimal network bandwidth allocation.

\begin{algorithm}[htb]
    \caption{Prefetching strategy}
    \label{alg.prefetch-caching}
    \KwIn{video stream $F_{i}$, lookup table $\{T_{i}\}_0^{R-1}$, client cache $C$}
    \KwOut{prefetched models $M_{p} = \{M_{i},\dots,,M_{j}\}$}
    \While{video stream}{
        Select the model $M_{i}$ for the current frame\;
        \label{alg.prefetch-caching.line2}
        $\{p_{i,0},p_{i,1},\dots,p_{i,R-1}\} \leftarrow$ the probability of transitioning from $M_{i}$ to $M_{j}$, $j \in \{0,1,\dots,r-1\}$\;
        \label{alg.prefetch-caching.line3}
        $M_{cache} \leftarrow M_{cache} \cup M_{N_{j}}$, with $M_{N_{j}} \leftarrow$ Top k models with highest transfer probabilities based $\{p_{i,0},p_{i,1},\dots,p_{i,R-1}\}$\;
        \label{alg.prefetch-caching.line4}
        Transmit $M_{p} \leftarrow M_{cache} \setminus C$ to the client\;
        \label{alg.prefetch-caching.line5}
    }
\end{algorithm}

The bandwidth allocation for the model stream is constrained, and the maximum bandwidth for downloading models is represented as $ \Delta_{bandwidth} = B_{hr} - B_{lr}$, where $B_{hr}$ and $B_{lr}$ denote the bitrate of high-resolution and low-resolution cloud game video streaming, respectively. The larger the disparity in bitrate before and after video stream compression, the more bandwidth can be set aside for downloading SR models. As per \cite{YouTube_encoding}, given that the original video is 1080p and the compressed video is 270p, a maximum of 7 Mbps bandwidth can be allocated for model transmission.

However, according to Algorithm \ref{alg.online_schedluer}, different frames may request different SR models, leading to a maximum model download bandwidth utilization of up to 40 Mbps. This is 6 times the maximum model bandwidth. Although employing an SR model to handle the entire video segment might deteriorate the enhancement quality, fortunately, due to the inherent similarity in the temporal aspect of video streams, when the current frame hits $M_{i}$, there is a high probability of encountering $M_{j}$, which is similar to $M_{i}$. Based on this observation, we propose a prefetching strategy to minimize the impact of model transmission on the latency of video streaming.

Algorithm \ref{alg.prefetch-caching} illustrates how \awRe~prefetches model weights to the client cache for video segments. We opt to send a set of SR models $\Omega = \{M_{i},\dots, M_{j}\}$ to the client simultaneously, where there is a high likelihood of $M_{j}$ being requested after $M_{i}$. We allow this series of models $\Omega = \{M_{i},\dots, M_{j}\}$ to process the video segment. For the current frame, we retrieve the optimal model $M_{i}$ from the lookup table using Algorithm~\ref{alg.online_schedluer}. For the model $M_{i}$, we compute the transfer matrix (\ie, line \ref{alg.prefetch-caching.line3}). We define the probability of transitioning from $M_{i}$ to $M_{j}$ as $p_{i,j}$, which is computed using \autoref{eq.transfer_matrix}. The higher the transition probability, the more similar the models are.

\begin{equation}
\centering
\begin{cases}
    p_{i,j} = softmax(d_{i,j}) \\
    d_{i,j} = \sum_{k=0}^{K-1}\max(d_{i,j}^k) \\
    d_{i,j}^k = \{S_c(\mu_i^k,\mu_j^{\hat{k}})|\hat{k} \in \{0,1,\dots,K-1\}\}\\
    j \in \{0,1,\dots,R-1\}
\end{cases}
\label{eq.transfer_matrix}
\end{equation} 

We select the top-k models based on the transfer matrix, considering the highest transfer probabilities (\ie, line \ref{alg.prefetch-caching.line4}). In addition, we maintain a model cache queue on the client side and use the Least Recently Used (LRU) algorithm, so that models already in the client cache need not be sent again (\ie, line \ref{alg.prefetch-caching.line5}).

\section{Implementation}\label{sec:impl}
We have implemented the server and client of \awRe~ on top of the state-of-the-art open-sourced framework, \emph{WebRTC}~\cite{WebRTC}. The algorithm processes for \awRe's server and client sides are respectively realized in \emph{Python} and \emph{C/C++}.
 
\noindent\textbf{Video Streaming Delivery.} We utilize \emph{FFmpeg}\cite{FFmpeg} (version 4.2.7) to encoder rendered video streaming with \emph{H.264}\cite{sullivan2005h264} preset \emph{veryslow} as the codec .To implement \awRe, we modify \emph{WebRTC} (version m85 4183) to obtain the actual timestamp of each frame, as well as the raw decoded and encoded frames generated by the \emph{H.264} codec.

\noindent\textbf{Training and Inference.} On the server side, we train the SR model using \emph{Pytorch} (version 1.13.1). On the client side, we utilize high-performance neural network inference computing frameworks, such as \emph{NCNN}\cite{NCNN} (version 20230517), to upscale frames received.

\section{Evaluation}\label{sec:evaluation}
In this section, we conduct a comprehensive evaluation of our method using open-source cloud gaming video datasets to demonstrate the advantages of our proposed methods.  Details about the video datasets and our evaluation are presented in Section \ref{sec.eval.exp_details}. The key findings of our experiments are summarized as follows:
\begin{itemize}
\item \awRe~reduces the training overhead by 44\%. By retrieving the appropriate SR model for new video streams, \awRe~attains an average PSNR improvement of 1.81 dB in overall video quality over the generic model without incurring additional training costs (\S\ref{sec.eval.compare_with_SOTA}).
\item \awRe~effectively strikes an effective balance between model download bandwidth and video bandwidth and ensure stable QoS (\S\ref{sec.eval.prefetch}).
\item \awRe~maintains low execution latency on the server side (averaging at 5.59 ms) and meets the real-time requirements for end-to-end latency on mobile devices (\S\ref{sec.eval.latency}).
\end{itemize}

\subsection{Experimental Setup}
\noindent\textbf{Testbed.}
To verify our system's performance and end-to-end latency, we implement \awRe~on the following testbed.
\begin{itemize}
    \item\noindent\textbf{Server-side.} Intel(R) Xeon(R) Gold 6330 CPUs with 512 GB Mem and a NVIDIA A100 (80GB) GPU.
    \item\noindent\textbf{Client-side.} Xiaomi 13 with Snapdragon 8 Gen2.
\end{itemize}

\label{sec.eval.exp_details}
\noindent\textbf{Datasets.} To validate the applicability of our algorithms, we conduct tests on real cloud gaming video datasets, \emph{GVSET}\cite{barman2018GVSET} and \emph{CGVDS}\cite{zadtootaghaj2020CGVDS}. These datasets comprise uncompressed raw gaming videos from distinct games, offering 1080p resolution videos for each game.
Following the YouTube recommendations \cite{YouTube_encoding}, we re-encode the raw gaming videos using the H.264 codec.The bitrates are respectively set to $\{500,2500,8000\}Kbps$, representing $\{270,540,1080\}P$ resolution videos at 30fps. 

It is important to note that because the algorithm is designed for an online problem, we split the videos in the dataset into a training set (the first half of the video) to fine-tune the SR model for video delivery and a validation set (the second half of the video) to evaluate the performance of image enhancement. The datasets are aligned with online cloud gaming but are quite different from offline video delivery, such as VoD, which allows training SR models on test data. Unless otherwise noted, we set 1080P as the target resolution. 



\noindent\textbf{Training Details.} According to \cite{khan2022survey,dao2022survey,lee2021survey}, each video from \emph{Part1} is segmented into sections of 10s duration. Our approach is applied to a variety of popular SR models, including the \emph{ultra-high} level of NAS\cite{yeo2018NAS}, WDSR\cite{yu2020wdsr} with 16 resblocks and EDSR\cite{lim2017edsr} with 16 resblocks. The LR video is divided into patches of $64\times64$ for a scaling factor of $2$ and $32\times32$ for a scaling factor of $4$ to accommodate the HR video size. We employ the Adam optimizer with $\beta_1$ = 0.9, $\beta_2$ = 0.999, $\epsilon$ = $10^{-8}$, and a batch size of 128. The L1 loss function is adopted. The learning rate is set as \num{2e-4} and undergoes decay at varying iterations based on the cosine function to $10^{-7}$. We train all neural networks on a single A100 (80GB) GPU.

\noindent\textbf{Patch Encoder.} For \awRe, we use a pre-trained \emph{Resnet18}\cite{resnet} on \emph{ImageNet}\cite{imagenet} to extract the embedding for each patch, with a dimensionality of 512. We use the function \emph{register\_forward\_hook} in \emph{Pytorch} to obtain the output of the average pooling layer as the embedding.


\subsection{Comparison with Baselines}
\label{sec.eval.compare_with_SOTA}
To demonstrate the quality improvement offered by our video delivery framework, we compare \awRe~with the state-of-the-art (SOTA) designs that have not undergone additional fine-tuning for validation set:

\begin{itemize}
    \item\noindent\textbf{Generic Super-resolution} uses a DNN SR model trained on the DIV2K benchmark dataset\cite{DIV2K}.
    \item\noindent\textbf{awDNN~\cite{yao2017awDNN}} proposes classifying fine-tuned models using fitted video labels. Since our algorithm does not use specific content labels, we categorize all videos into a single group for fairness. This means that we use one SR model to fit all the data via fine-tuning. 
     \item\noindent\textbf{\randomRe} involves randomly reusing the SR model from the model pool while maintaining all other algorithms as in \awRe~.
\end{itemize}

\noindent\textbf{Videos selection and fine-tuning models.}
We assume that the video stream of the training set arrives in a random chronological order. We perform online scheduling of the video stream based on Algorithm \ref{alg.online_schedluer}, where $\lambda$ is set to 10, $\beta$ is set to 0.8, and $\alpha$ is set to 0.65. This process includes selecting the reuse model and determining the necessity for additional fine-tuning. In our implementation, we treat video segments as the smallest training unit. If the number of frames in a video segment requiring retraining exceeds the threshold set at 0.65, we employ an independent model to fit that segment. We update the lookup table according to Algorithm \ref{alg.model_encoder} and set the number of cluster centers to 5. Finally, \awRe~conducts additional fine-tuning for 20 of the 36 video segments, resulting in a 44\% reduction in additional training cost, as detailed in \autoref{tbl.look-up_table}. The first column lists various cloud gaming video streams, while columns 2 to 4 represent the sizes of individual 10-second video segments for each stream. A checkmark indicates fine-tuning for that video segment, while a cross signifies that no adjustment is needed, as similar video segments have already been fine-tuned. We observe that fine-tuning is only applied to the first video segment for games with relatively stable visuals, such as FIFA17 and LoL, while in games with more significant visual changes, such as H1Z1 and PU, fine-tuning is frequently required for each video segment.
\begin{table}[htbp]
    \centering
    \caption{Fine-tuned Video segments of cloud game according to \awRe~}
    \label{tbl.look-up_table}
    \begin{tabular}{|c|c|c|c|}
        \hline
        \multirow{2}{*}{Video Streaming} & \multicolumn{3}{c|}{Time Intervals} \\ \cline{2-4}
         & 0-10s & 10s-20s & 20s-30s \\
        \hline
        CSGO / DiabloIII & \multirow{4}{*}{$\checkmark$} & \multirow{4}{*}{$\times$} & \multirow{4}{*}{$\times$} \\
        Dota2 / FIFA17 & ~ & ~ &\\
        LoL / StarCraftII & ~ & ~ & \\
        Hearthstone & ~ & ~ & \\
        \hline
        H1Z1 / ProjectCars & $\checkmark$ & $\times$ & $\checkmark$ \\
        \hline
        Heroes / PU / WoW & $\checkmark$ & $\checkmark$ & $\checkmark$ \\
        \hline
    \end{tabular}
\end{table}

On the validation set, we only execute the retrieving part of Algorithm \ref{alg.online_schedluer} (\eg, lines \ref{alg.online_schedluer.2}-\ref{alg.online_schedluer.13}). We assume that all models retrieved by \awRe~are available in the client's cache. \autoref{tbl.retrival_psnr} compares our algorithm with state-of-the-arts.

\noindent{\textbf{PSNR improvements over baselines.} Our approach shows significant improvement of PSNR by an average of 1.81 dB, compared against generic SR techniques. In contrast to awDNN\cite{yao2017awDNN}, which uses a single model for all videos, \awRe~permits the use of different models for different video streams, ensuring a more effective fit. Although awDNN may perform well on some video streams with substantial variations (such as CSGO and ProjectCars), \awRe~still achieves PSNR improvements ranging from 0.12 dB to 2.28 dB on most video streams. This advantage of \awRe~reflects the inherent trade-off between generalization and specialization in any machine learning method. As the "No Free Lunch" theorem \cite{wolpert1996lack} suggests, an increase in model coverage may compromise its performance. Unlike the broad generalization seen with a universal model, \awRe~allows the selection of an SR model from the lookup table, one that has shown success on a similar video stream, thereby achieving better specialization. When an SR model is adapted to fit diverse data, it may lose specific information and gradually become more generalized, as seen with awDNN. Conversely, if SR models are randomly chosen for reuse, as in the case of \randomRe, the quality might even be inferior to that of a generic model, since model's specialization does not align with the inference data. This discrepancy clearly underscores the effectiveness of \awRe~in selecting the appropriate model for each frame.

\begin{table*}[htbp]
\centering
\caption{Comparison of PSNR between content-aware retrieval (\awRe) and the State-of-the-Art methods.}
\begin{tabular}{cccccccccccccc}
\toprule
\multirow{2}{*}{Model} & \multirow{2}{*}{Method} & \multicolumn{2}{c}{CSGO} & \multicolumn{2}{c}{DiabloIII} & \multicolumn{2}{c}{Dota2} & \multicolumn{2}{c}{FIFA17} & \multicolumn{2}{c}{H1Z1} & \multicolumn{2}{c}{Hearthstone} \\
~ & ~ & x2 & x4 & x2 & x4 & x2 & x4 & x2 & x4 & x2 & x4 & x2 & x4\\
\midrule
\multirow{4}{*}{NAS} & Generic & 36.90 & 27.49 & 34.30 & 28.42 & 34.35 & 27.41 & 35.12 & 27.02 & 38.24 & 29.45 & 33.53 & 25.86\\
~ & awDNN~\cite{yao2017awDNN} & \pmb{37.62} & \pmb{27.79} & 35.34 & 28.85 & 35.50 & 28.11 & 35.93 & 27.54 & \pmb{38.86} & 29.89 & 34.02 & 26.92 \\
~ & \randomRe & 37.13 & 27.19 & 34.91 & 28.29 & 34.82 & 26.94 & 35.23 & 25.97 & 38.21 & 28.89 & 33.55 & 25.11 \\ 
~ & \awRe~ & 37.52 & 27.77 & \pmb{36.84} & \pmb{29.36} & \pmb{35.95} & \pmb{28.74} & \pmb{36.19} & \pmb{27.76} & 38.83 & \pmb{30.33} & \pmb{34.36} & \pmb{27.49} \\

\hline
\multirow{4}{*}{WDSR} & Generic & 38.45 & 27.46 & 35.74 & 28.34 & 35.57 & 27.20 & 35.82 & 26.80 & 40.03 & 29.47 & 34.58 & 25.47\\
~ & awDNN~\cite{yao2017awDNN} & \pmb{39.62} & \pmb{27.97} & 38.67 & 29.06 & 37.62 & 28.30 & 36.86 & 27.66 & 41.38 & 30.25 & 35.71 & 27.10 \\
~ & \randomRe & 39.15 & 27.45 & 37.23 & 28.62 & 36.46 & 27.38 & 36.35 & 26.64 & 40.60 & 29.33 & 35.15 & 25.85 \\
~ & \awRe~ & 39.56 & 27.87 & \pmb{40.01} & \pmb{29.51} & \pmb{38.35} & \pmb{28.90} & \pmb{37.14} & \pmb{27.94} & \pmb{41.48} & \pmb{30.58} & \pmb{36.10} & \pmb{27.77} \\
 
\hline
\multirow{4}{*}{EDSR} & Generic & 38.36 & 27.51 & 35.55 & 28.44 & 35.44 & 27.35 & 35.74 & 26.92 & 39.92 & 29.58 & 34.45 & 25.74\\
~ & awDNN~\cite{yao2017awDNN} & \pmb{39.68} & \pmb{28.05} & 39.76 & 29.33 & 38.08 & 28.62 & 36.97 & 27.95 & 41.38 & 30.55 & 35.72 & 27.38 \\
~ & \randomRe & 38.74 & 27.20 & 37.40 & 28.30 & 36.07 & 26.98 & 36.14 & 26.08 & 40.35 & 28.65 & 34.89 & 25.21 \\
~ & \awRe~ & 39.62 & 27.76 & \pmb{40.78} & \pmb{29.84} & \pmb{38.85} & \pmb{29.20} & \pmb{37.33} & \pmb{28.10} & \pmb{41.66} & \pmb{30.57} & \pmb{36.02} & \pmb{27.75} \\
\bottomrule
\\
\toprule
~ & ~ & \multicolumn{2}{c}{Heroes} & \multicolumn{2}{c}{LoL} & \multicolumn{2}{c}{ProjectCars} & \multicolumn{2}{c}{PU} & \multicolumn{2}{c}{StarCraftII} & \multicolumn{2}{c}{WoW} \\
~ & ~ & x2 & x4 & x2 & x4 & x2 & x4 & x2 & x4 & x2 & x4 & x2 & x4 \\
\midrule
\multirow{4}{*}{NAS} & Generic & 31.96 & 24.44 & 33.68 & 25.72 & 37.23 & 29.10 & 34.33 & 25.06 & 31.78 & 26.24 & 31.62 & 26.95\\
~ & awDNN~\cite{yao2017awDNN} & 33.08 & 25.36 & 35.08 & 27.47 & \pmb{37.79} & \pmb{29.45} & 35.54 & 25.30 & 32.57 & 26.69 & 32.53 & 27.55 \\
~ & \randomRe & 32.15 & 23.91 & 34.01 & 25.77 & 37.06 & 28.58 & 34.73 & 24.73 & 31.87 & 25.67 & 31.83 & 26.44 \\ 
~ & \awRe~ & \pmb{34.07} & \pmb{26.46} & \pmb{36.05} & \pmb{28.16} & 37.64 & 29.30 & \pmb{36.30} & \pmb{25.44} & \pmb{33.05} & \pmb{27.15} & \pmb{34.21} & \pmb{28.27} \\

\hline
\multirow{4}{*}{WDSR} & Generic & 33.23 & 24.27 & 35.17 & 25.50 & 38.41 & 29.04 & 36.18 & 25.02 & 33.08 & 26.08 & 32.72 & 26.82\\
~ & awDNN~\cite{yao2017awDNN} & 35.60 & 25.62 & 37.62 & 27.64 & 39.29 & \pmb{29.77} & 37.80 & 25.42 & 34.97 & 26.96 & 35.31 & 27.72 \\
~ & \randomRe & 34.19 & 24.71 & 36.16 & 25.83 & 38.72 & 28.99 & 36.71 & 25.01 & 33.59 & 26.00 & 33.47 & 26.86 \\
~ & \awRe~ & \pmb{36.42} &  \pmb{26.68} & \pmb{38.45} & \pmb{28.33} & \pmb{39.39} & 29.60 & \pmb{38.52} & \pmb{25.52} & \pmb{35.23} & \pmb{27.51} & \pmb{37.10} & \pmb{28.42} \\

\hline
\multirow{4}{*}{EDSR} & Generic & 32.93 & 24.40 & 34.96 & 25.63 & 38.27 & 29.12 & 35.79 & 25.05 & 32.85 & 26.22 & 32.52 & 26.90\\
~ & awDNN~\cite{yao2017awDNN} & 36.05 & 26.16 & 38.05 & 28.04 & 39.21 & \pmb{29.72} & 38.00 & 25.47 & 35.53 & 27.19 & 36.55 & 28.04 \\
~ & \randomRe & 33.71 & 24.26 & 35.72 & 25.35 & 38.06 & 28.50 & 36.63 & 24.64 & 33.23 & 25.61 & 32.91 & 26.54 \\
~ & \awRe~ & \pmb{36.94} & \pmb{27.05} & \pmb{39.35} & \pmb{28.73} & \pmb{39.27} & 29.43 & \pmb{38.86} & \pmb{25.41} & \pmb{35.70} & \pmb{27.52} & \pmb{39.14} & \pmb{28.96} \\
\bottomrule
\end{tabular}
\label{tbl.retrival_psnr}
\end{table*}

\subsection{Prefetch v.s. No-prefetch}
\label{sec.eval.prefetch}
As discussed in Section \ref{sec.sys.prefetching_cache}, we must consider the bandwidth consumed by downloading the SR model in real-world scenarios. It's impractical to download every model that \awRe~recommends for each frame. However, unfulfilled model download requests can result in a decline in QoS. To address this, we design a prefetching strategy. We then conduct experiments to verify whether our prefetching strategy can enhance the model hit ratio and reduce the latency of activating SR models in real-time generated contents.


\begin{itemize}
    \item\noindent\textbf{No-prefetch} transmits only one model for the current frame to the client as retrieved by Algorithm \ref{alg.online_schedluer}.
    \item\noindent\textbf{Prefetch} determines the prefetched models based on Algorithm \ref{alg.prefetch-caching}, prefetching the top three models with the highest transition probability to the client.
\end{itemize}
\noindent\textbf{Set-up for evaluating.} Note that Prefetching may transmit three models simultaneously, while No-prefetching only transmits the SR model needed for the current frame. To ensure equal bandwidth consumption, we set Prefetching to execute every 30 seconds, while No-prefetching operates every 10 seconds. Assuming the client's model buffer size is three, we employ the Least Recently Used (LRU) algorithm for cache replacement. The generic model is used if the required model is not in the cache. 

\begin{figure}[htbp]
    \centering
    \begin{minipage}[c]{0.48\linewidth}
        \centering
        \includegraphics[width=\linewidth]{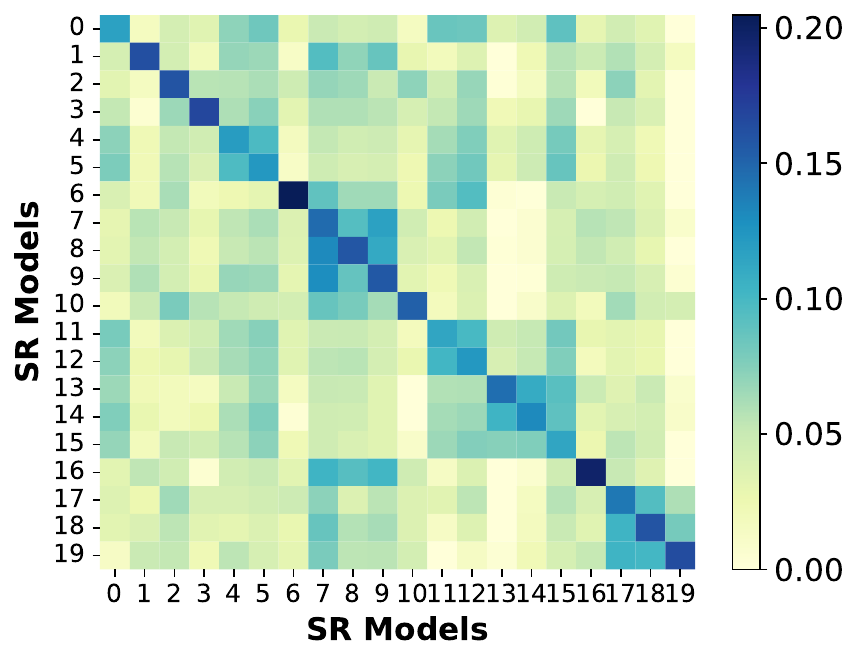}
        \subcaption{Transfer Matrix.}
        \label{fig.prefetch_strategy.transfer_matrix}
    \end{minipage}
    \begin{minipage}[c]{0.48\linewidth}
        \centering
        \includegraphics[width=\linewidth]{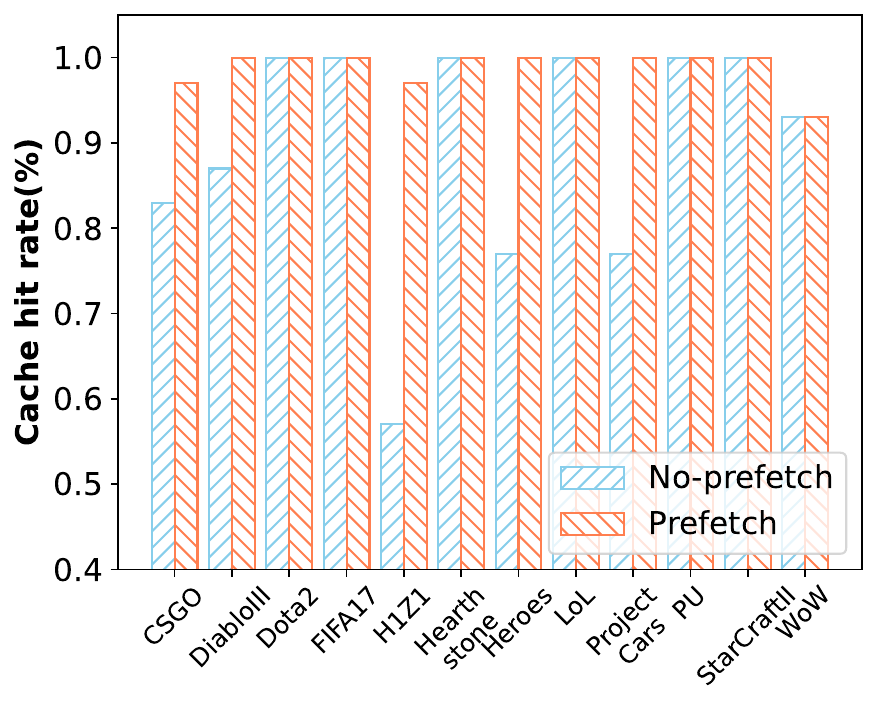}
        \subcaption{Cache Hit Ratio.}
        \label{fig.prefetch_strategy.cache_hit_rate}
    \end{minipage}  
    \\
    \begin{minipage}[c]{0.48\linewidth}
        \centering
        \includegraphics[width=\linewidth]{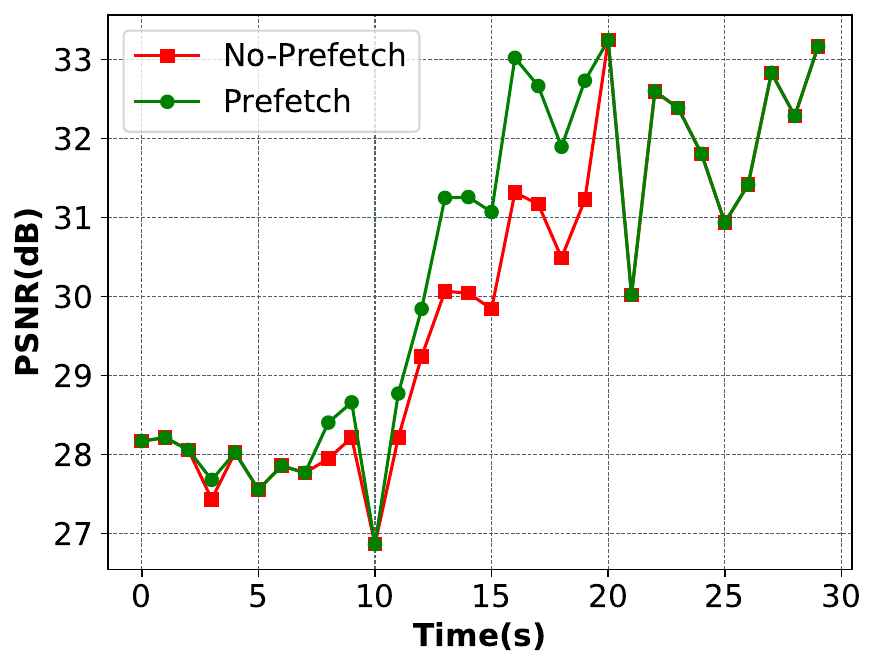}
        \subcaption{PSNR of H1Z1.}
        \label{fig.prefetch_strategy.prefetch.H1Z1}
    \end{minipage}
    \begin{minipage}[c]{0.48\linewidth}
        \centering
        \includegraphics[width=\textwidth]{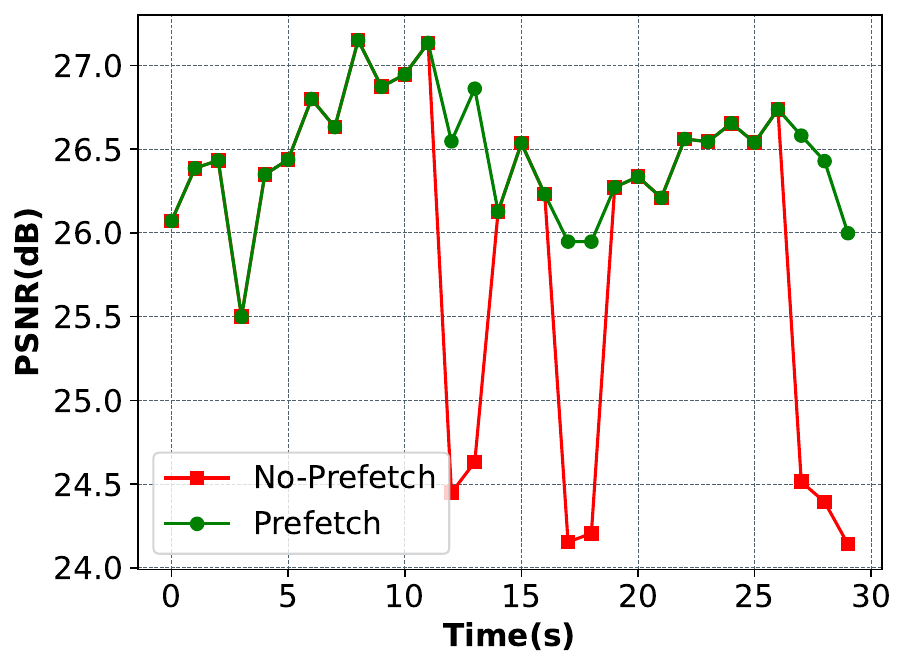}
        \subcaption{PSNR of Heroes.}
        \label{fig.prefetch_strategy.prefetch.Heroes}
    \end{minipage}
    \caption{(a) Visualization results of the transfer matrix that is defined by \autoref{eq.transfer_matrix}. (b-d) Comparison of the hit ratio and PSNR between Prefetch and No-prefetch.}
\label{fig.prefetch_strategy}
\end{figure}

\autoref{fig.prefetch_strategy.transfer_matrix} visualizes the results of the transfer matrix. The x-axis and y-axis represent the SR models fine-tuned according to \awRe, as shown in \autoref{tbl.look-up_table}. Each cell represents the probability of transitioning from the model $M_{i}$ to the model $M_{j}$. The darker the color, the higher the probability. This variation in transition probabilities reflects the different similarities between video segments. The highest probability represents the transition to the same model, given that the prefetched models must include the optimal choice of the current frame.

\autoref{fig.prefetch_strategy.cache_hit_rate} displays the cache hit ratio for the 12 games of the validation set. Prefetching's ratio is in red, while the ratio without prefetching is in blue. Our Prefetch has achieved a 100\% cache hit rate on most games. The No-prefetching strategy, however, have significant cache misses, especially in the games with rapid scene changes, such as H1Z1 and Heroes. As illustrated in \autoref{fig.prefetch_strategy.prefetch.H1Z1} and \autoref{fig.prefetch_strategy.prefetch.Heroes}, the green line represents the PSNR with our prefetching strategy, and the red line represents that without prefetching. The higher cache hit ratio from Prefetching, which is achieved with almost comparable bandwidth usage, result in better PSNR. If \awRe~fails to find the required model within the cache, leading to the use of a generic model, as in No-prefetching, it would result in a noticeable drop in image enhancement quality.
In summary, our prefetching strategy enhances the hit ratio of the SR model in clients and stabilizes the Quality of Service (QoS).

\subsection{Deployment latency}
\label{sec.eval.latency}

The end-to-end latency of the system consists of two parts: the latency on the server side, resulting from performing \awRe's algorithms, and client-side latency, which is associated with the execution of image enhancement.

\begin{figure}[htb]
    \centering
    \includegraphics[width=0.75\linewidth]{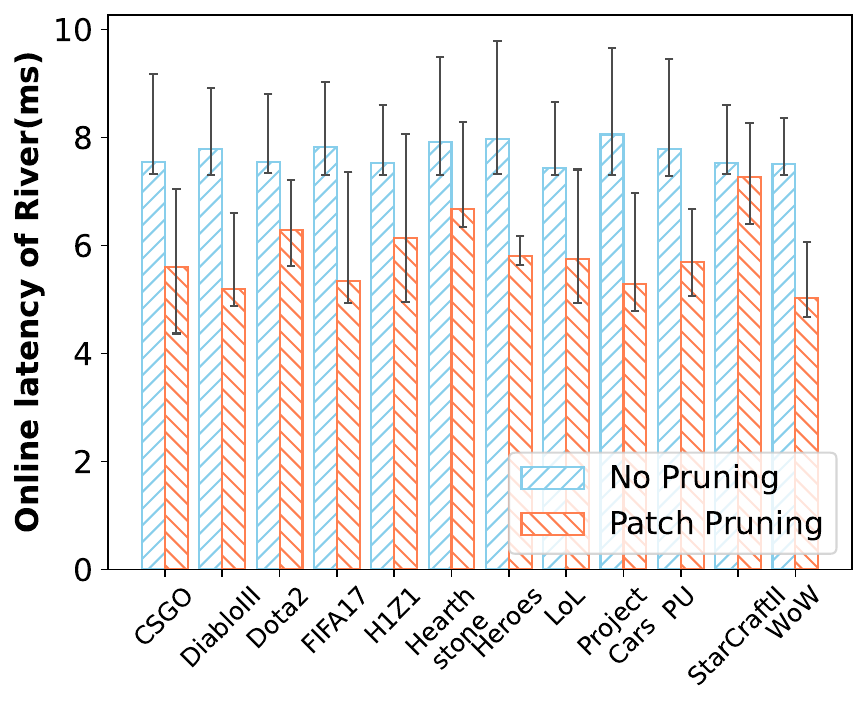} 
    \caption{Online latency of \awRe~on video games.} 
    \Description{Online latency of \awRe~on video games.} 
    \label{fig.alg_latency} 
\end{figure}

\noindent\textbf{Server-side.} While \awRe~reduces the model fine-tuning delay, our method does introduce latency due to the algorithm, compared to directly executing a generic model on the client side. Since Algorithm \ref{alg.model_encoder} and the transfer matrix $\{p_{i,0},p_{i,1},...,p_{i,r-1}\}$ of Algorithm \ref{alg.prefetch-caching} can be performed offline, the online latency of \awRe~is almost determined by Algorithm \ref{alg.online_schedluer}. We then implement the algorithm on the server and validate the latency of running our algorithms, with \autoref{fig.alg_latency} showing its average for different games. As detailed in Section \ref{sec.sys.scheduler}, we employ patch pruning to minimize the computational effort of Algorithm \ref{alg.online_schedluer} during online operations.  

The latency with pruned patches, represented in red, is compared to the un-pruned latency, marked in blue. The result suggests pruning can reduce the execution delay by approximately 25\% compared to computing all patches, largely related to the amount of useful information in the frames. \eg, the percentage of patches with high edge scores in the frame. The average execution latency of \awRe~ on the server side is about 5.59 ms.

\noindent\textbf{Client-side.} To better reflect the real-world application scenarios of cloud gaming, 
we select mobile devices as client systems and implement the SR model using 16-bit floating-point precision (\emph{FP16}).
We set the target resolution as 1280x720 and define real-time as end-to-end latency less than 50 ms on mobile devices\cite{jo2021SRlut,ignatov2021real,zhan2021achieving,xu2022litereconfig}. We deploy the SR model on the Xiaomi 13 with Snapdragon 8 Gen2 and utilize \emph{NCNN}\cite{NCNN} for inference computing with setting NAS\cite{yeo2018NAS} as the SR backbone. Besides, we propose a rearrangement operator to accelerate the computation for \awRe, \eg, rearranging the feature dimension from $(c,h,w)$ to $(c*r*r,h/r,w/r)$. This operation can reduce the input size and maximize the convolution's parallelism. The end-to-end latency and PSNR between the generic model and \awRe~are shown in \autoref{fig.mobile_latency}.


\begin{figure}[htb]
    \begin{minipage}[c]{0.48\linewidth}
        \centering
        \includegraphics[width=\linewidth]{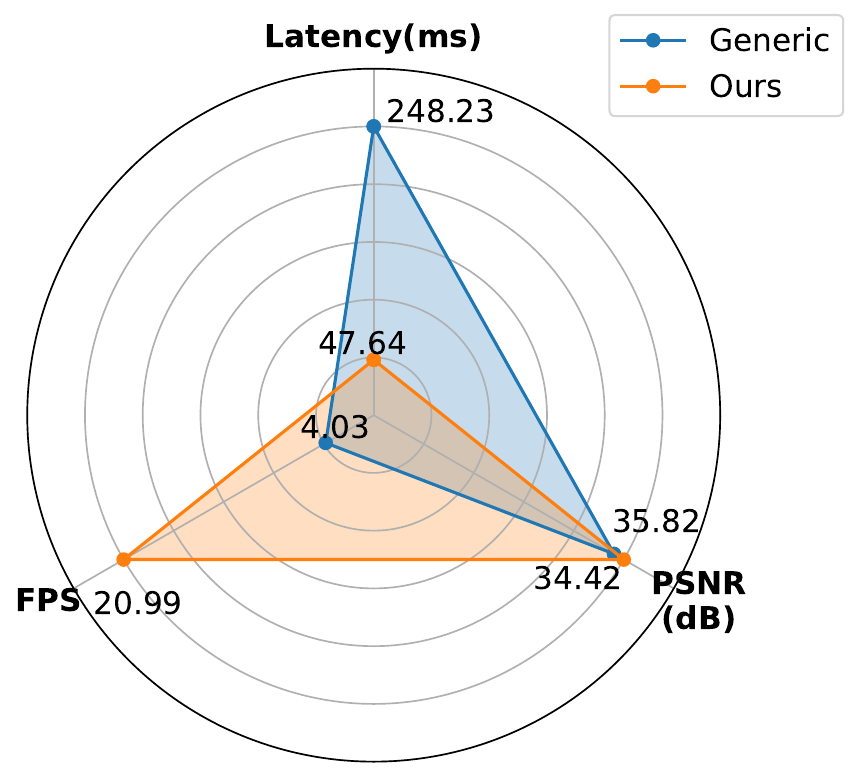}
        \subcaption{Scaling factor 2.}
        \label{fig.mobile_latency.x2}
    \end{minipage} 
    \begin{minipage}[c]{0.48\linewidth}
        \centering
        \includegraphics[width=\linewidth]{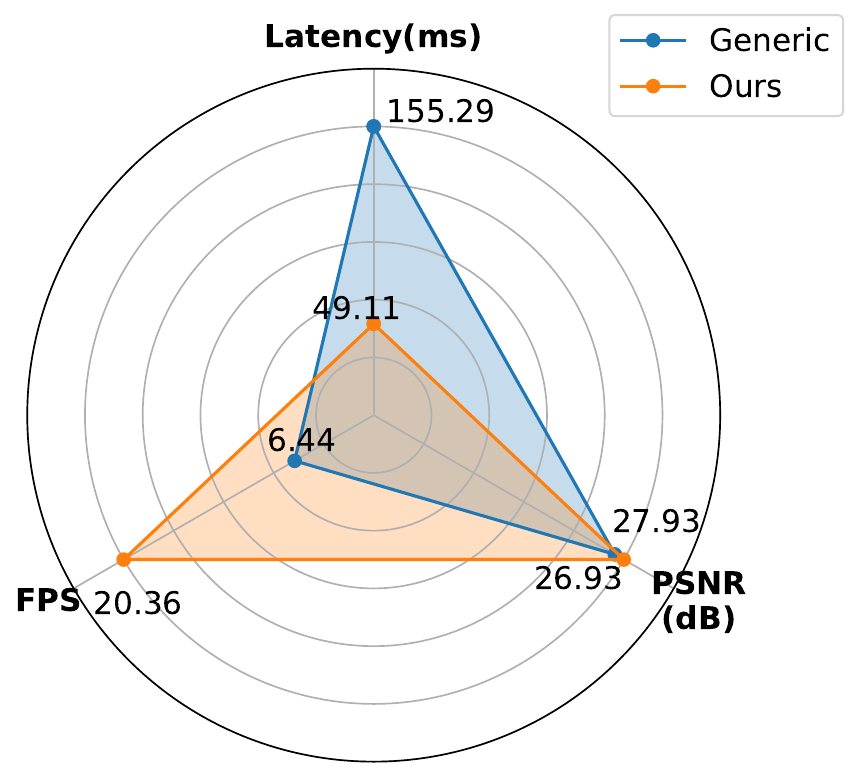}
        \subcaption{Scaling factor 4.}
        \label{fig.mobile_latency.x4}
    \end{minipage}
    \caption{End-to-end latency and PSNR between ours and generic super-resolution on Xiaomi 13.}
    \Description{End-to-end latency and PSNR between ours and generic super-resolution on Xiaomi 13.}
    \label{fig.mobile_latency}
\end{figure} 

The blue represents the performance of the generic model, and the orange represents \awRe. The result shows our method clearly significantly outperforms the generic model in both end-to-end latency and PSNR. Although the rearrangement operator can facilitate up to 5x speed-up in computation to meets real-time requirements, it concurrently incurs a certain degradation in model performance due to the reduction in the receptive field. Nonetheless, thanks to the ability of \awRe~to select the appropriate model based on the content of the current frame, we still maintain a lead over the benchmark by 1.2dB in terms of PSNR. 


\subsection{Ablation Study}
In this part, we provide further insights into the rationale behind the setting used in the previous sections.
\subsubsection{Frame or patch embedding}
In Algorithm \ref{alg.model_encoder} and \ref{alg.online_schedluer}, we divide frames into patches rather than computing features per frame. In this ablation study, we compare the impact of patch and frame-level features on the performance of \awRe. We chose NAS\cite{yeo2018NAS}, WDSR\cite{yu2020wdsr}, and EDSR\cite{lim2017edsr} as the super-resolution model. The upscaling factor is set to 4. Aside from the difference in the unit for the algorithms, other parameters are kept consistent with those in Section \ref{sec.eval.compare_with_SOTA}.

\begin{table}[htb]
\centering
\caption{Comparison of PSNR with various units of \awRe.}
\resizebox{\linewidth}{!}{
\begin{tabular}{ccccccc}
\toprule
Model & \multicolumn{2}{c}{Method} & Heroes & ProjectCars & PU & WoW \\
\midrule
\multirow{3}{*}{NAS} & \multicolumn{2}{c}{Generic} & 24.44 & \color{blue}29.10 & \color{blue}25.06 & 26.24 \\
\cline{2-3}
~ &\multirow{2}{*}{\awRe} & Frame & \color{blue}26.37 & 28.81 & 24.82 & \color{blue}27.62\\
~ & ~ &  Patch & \color{red}26.46 & \color{red}29.30 & \color{red}25.44 & \color{red}28.26\\
\hline
\multirow{3}{*}{WDSR} & \multicolumn{2}{c}{Generic} & 24.27 & 29.04 & 25.02 & 26.82 \\
\cline{2-3}
~ &\multirow{2}{*}{\awRe} &  Frame & \color{blue}26.59 & \color{blue}29.26 & \color{blue}25.06 & \color{blue}27.84\\
~ & ~ & Patch & \color{red}26.68 & \color{red}29.60 & \color{red}25.52 & \color{red}28.42\\
\hline
\multirow{3}{*}{EDSR} & \multicolumn{2}{c}{Generic} & 24.40 & \color{blue}29.12 & \color{blue}25.05 & 26.90 \\
\cline{2-3}
~ &\multirow{2}{*}{\awRe} & Frame & \color{blue}26.98 & 28.69 & 24.75 & \color{blue}28.05\\
~ & ~ &  Patch & \color{red}27.05 & \color{red}29.43 & \color{red}25.41 & \color{red}28.96\\
\bottomrule
\end{tabular}}
\label{tbl.frame_embedding}
\end{table}
The results of this comparison are shown in \autoref{tbl.frame_embedding}, with the best outcomes in red and the second-best in blue. These findings indicate that calculating based on frame features performs significantly worse than those based on patches. The enhancement quality is even lower than the generic model when using frame-level features for the algorithms. This is due to the high diversity in cloud gaming and the finer granularity offered by patches compared to frames, leading to more precise feature extraction. The approach of dividing frames into patches is similar to using blocks in video compression, such as in the \emph{H.264}\cite{sullivan2005h264} codec. This comparison demonstrates the importance of patch-level calculations in improving the performance of \awRe.

\subsubsection{The importance of different patches}
\label{sec.patch_pruning}

\begin{table*}[htb]
\centering
\caption{Comparison of PSNR with different data overfitting methods.}
\begin{tabular}{ccccccccc}
\toprule
\multicolumn{2}{c}{Method} & CSGO & DiabloIII & Dota2 & FIFA17 & H1Z1 & Hearthstone \\
\midrule
\multicolumn{2}{c}{Generic} & 27.20 & 28.87 & 27.80 & 25.14 & 28.94 & 27.45\\
\cline{1-2}
\multirow{4}{*}{Overfitting} & All patch & \color{red}{28.11}  & \color{red}{31.68} & \color{red}{30.07} & \color{red}{26.33} & \color{red}{30.54} & \color{red}{30.63}\\
~ & Patch-pruning  & \color{blue}{28.01} & \color{blue}{31.42} & \color{blue}{30.01} & \color{blue}{26.25} & \color{blue}{30.34} & \color{blue}{30.56}\\
\cline{2-2}
~ & $\triangle_{PSNR}$ & 0.10 & 0.26 & 0.06 & 0.08 & 0.20 & 0.07\\
~ & $\triangle_{patch}$ & 7313/17000 & 4797/17000 & 9070/17000 & 7716/17000 & 3865/17000 & 11506/17000\\
\bottomrule
\\
\toprule
~ & ~ &  HeroesOfTheStorm & LoL & ProjectCars & PU & StarCraftII & WoW \\
\midrule
\multicolumn{2}{c}{Generic} & 24.53  & 25.84  & 27.25 & 29.97 & 28.45 & 25.16\\
\cline{1-2}
\multirow{4}{*}{Overfitting} & All patch & \color{red}{27.41}  & \color{red}{29.18} & \color{red}{28.63} & \color{red}{33.54} & \color{red}{30.28} & \color{red}{26.59}\\
~ & Patch-pruning & \color{blue}{27.27} & \color{blue}{29.09} & \color{blue}{28.52} & \color{blue}{33.34} & \color{blue}{30.19} & \color{blue}{26.41}\\
\cline{2-2}
~ & $\triangle_{PSNR}$ & 0.14 & 0.09 & 0.11 & 0.20 & 0.09 & 0.18\\
~ & $\triangle_{patch}$ & 9928/17000 & 5953/17000 & 8727/17000 & 3258/17000 & 11752/17000 & 7078/17000 \\
\bottomrule
\end{tabular}
\label{tbl.patch_pruning}
\end{table*}

We study whether data filtering would result in a loss of training effectiveness for super-resolution. NAS\cite{yeo2018NAS} is selected as the SR model with an upscaling factor set to 4. For training and validation, we utilize the \emph{GVSET}\cite{barman2018GVSET} dataset and we divide the LR frame into multiple $32\times32$ patches for training. 
We compared two data overfitting methods. The first method fine-tunes only those patches with edge scores above a threshold $\lambda$, 10, indicating higher complexity with these patches. 
The second approach trains the model using all available data. We use a generic SR model trained on DIV2K\cite{DIV2K} as the baseline.

As shown in \autoref{tbl.patch_pruning}, the red represents the best PSNR, and the blue represents the second-best. We use $\triangle_{PSNR}$ to represent the difference in PSNR between the two methods and $\triangle_{patch}$ to indicate the change in data volume before and after pruning. Even though pruning the data leads to a performance drop of approximately 0.1dB, it is still significantly better than the baseline, with an average improvement of around 2.06 dB. 
Our experiments demonstrate that by judicious selection of training images, it is possible to reduce the data by 50\% with slight performance degradation. It is critical to reduce the additional training burden, 

\subsubsection{Different k for lookup table}

We further investigate the impact of the number of clusters in Algorithm \ref{alg.model_encoder} on the performance of \awRe. We experiment with different values of $K$ to generate lookup tables and compare the performance of retrieval models across the 12 games in \emph{GVSET}\cite{barman2018GVSET}. We use NAS\cite{yeo2018NAS} as the backbone with an upscaling factor of 4. 

\begin{figure}[htb]
  \centering
  \begin{minipage}[c]{0.48\linewidth}
    \centering
    \includegraphics[width=\linewidth]{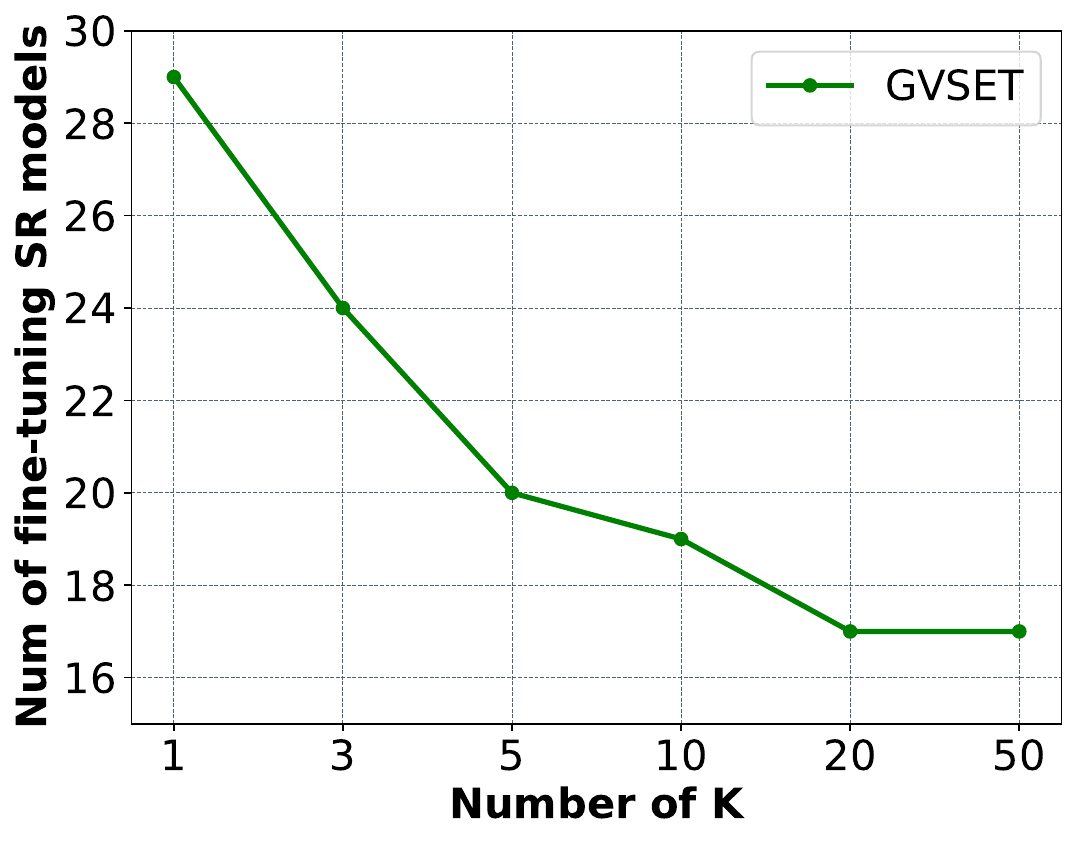}
    \subcaption{Num of fine-tuning models.}
    \label{fig.dif_k.num}
  \end{minipage}
  \begin{minipage}[c]{0.48\linewidth}
    \centering
    \includegraphics[width=\linewidth]{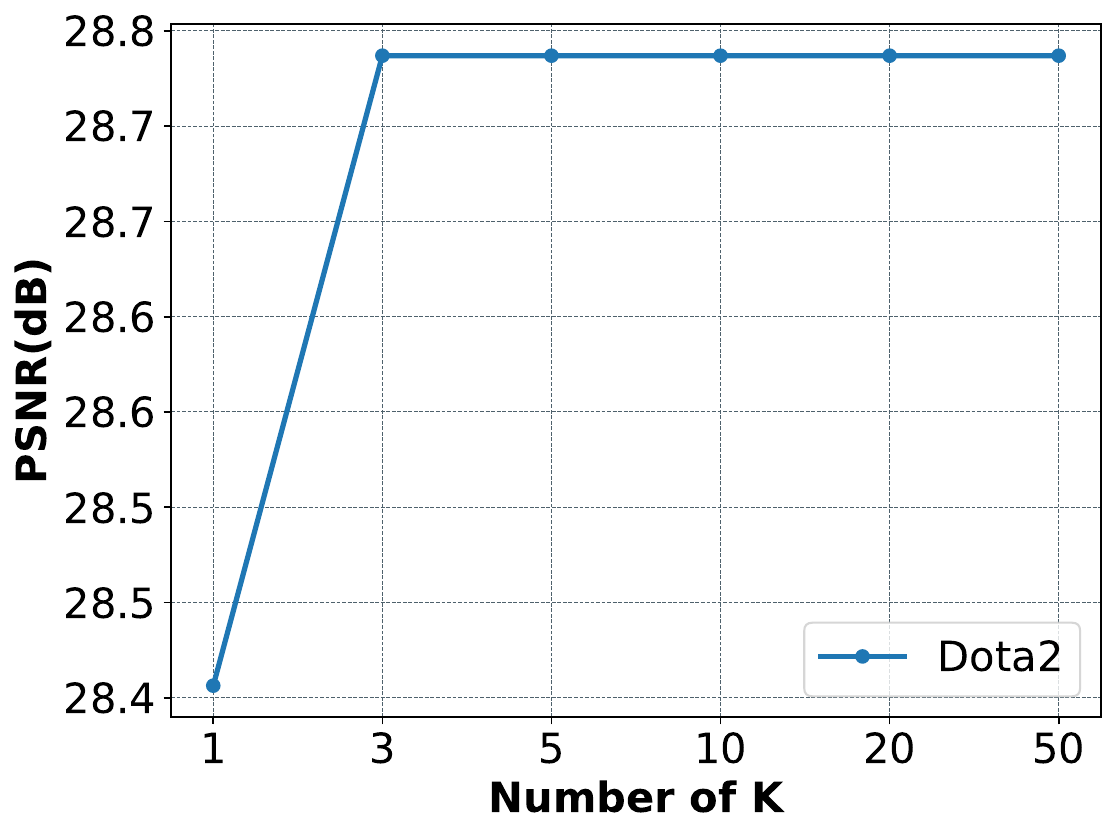}
    \subcaption{PSNR of Dota2.}
    \label{fig.dif_k.dota2}
  \end{minipage}  
  \\
  \begin{minipage}[c]{0.48\linewidth}
    \centering
    \includegraphics[width=\linewidth]{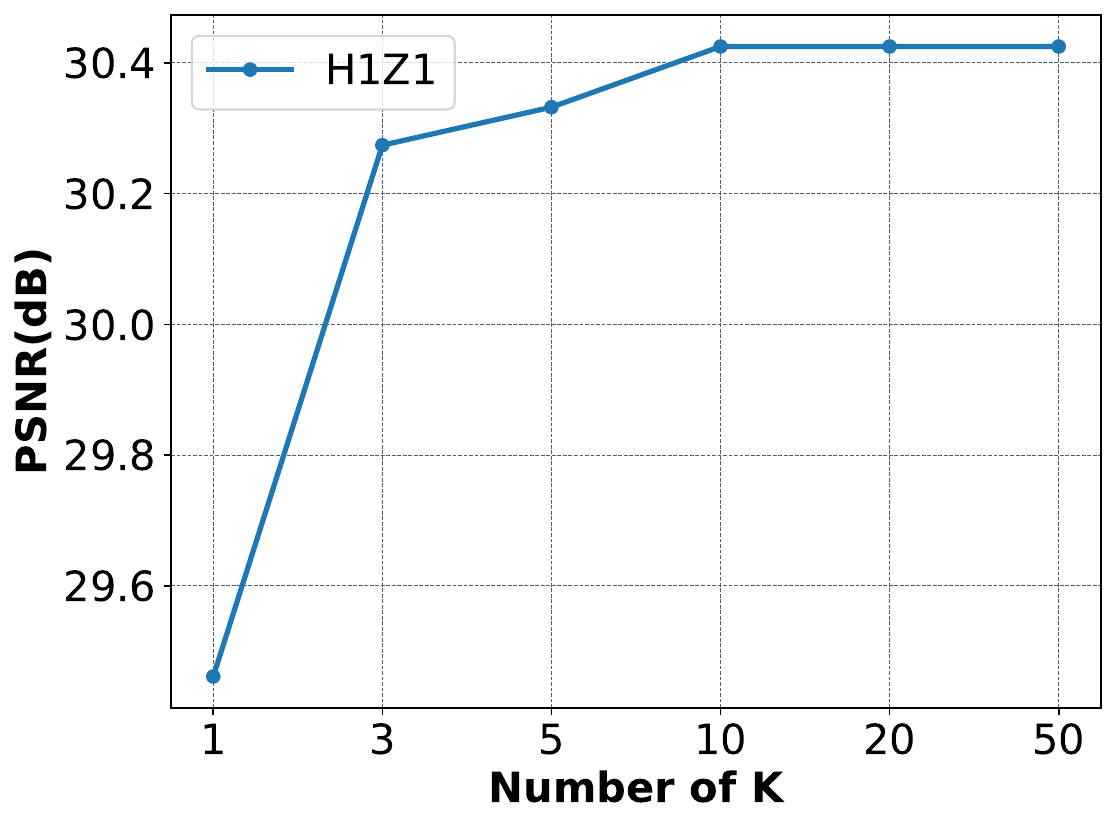}
    \subcaption{PSNR of H1Z1.}
    \label{fig.dif_k.h1z1}
  \end{minipage}
  \begin{minipage}[c]{0.48\linewidth}
    \centering
    \includegraphics[width=\linewidth]{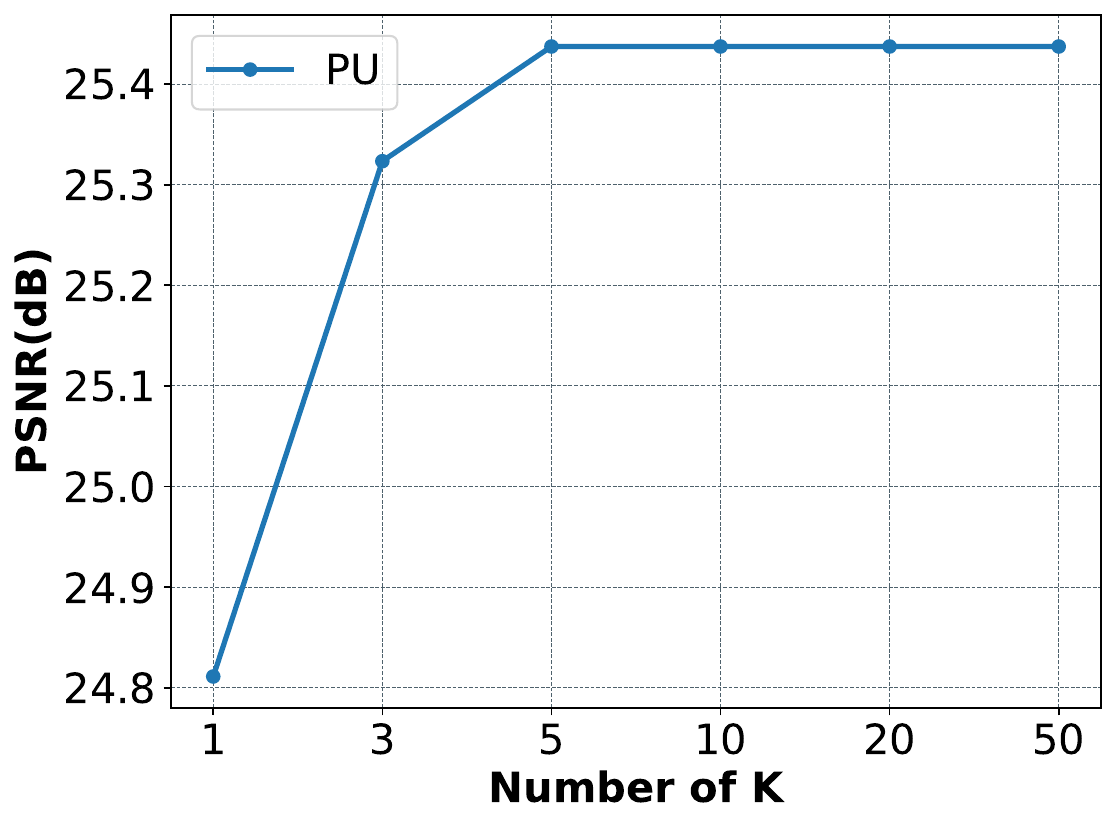}
    \subcaption{PSNR of PU.}
    \label{fig.dif_k.pu}
  \end{minipage}
  \caption{(a) Illustrates the number of fine-tuned models varying $k$. (b)-(d) Show the performance of \awRe~on different games with different values of $k$.}
  \Description{(a) Illustrates the number of fine-tuned models varying $k$. (b)-(d) Show the performance of \awRe~on different games with different values of $k$.}
  \label{fig.dif_k}
\end{figure}

As shown in \autoref{fig.dif_k.num}, the number of video segments requiring fine-tuning decreases as $K$ increases.
The performance of the retrieval models improves when $K$ increases progressively and stabilizes after $K$ reaches 5. The increase in $K$ results in more precise cluster centers, thereby leading to the more precise encoding of SR models according to Algorithm \ref{alg.model_encoder} and improving model retrieval accuracy. However, increasing $K$ does not continuously yield benefits, as a larger $K$ implies a need for more storage space and a higher retrieval overhead for the lookup table. The time to query the lookup table is about 1 ms when k is 5 and about 7 ms when k is 50. Therefore, choosing an appropriate $K$ (approximately 5) for clustering is the key to balance the overhead of training and model retrieval.

\section{Conclusion}
 In this paper, we propose \awRe, a novel online cloud gaming delivery framework based on content-aware retrieval to enhance the video quality efficiently. Addressing the limitations of neural-enhanced video delivery, \awRe~avoids the need for independent model fine-tuning for each video chunk. To achieve high efficiency, it leverages three novel design components: a content-aware encoder that encodes the SR model and manages the model lookup table, an online scheduler for retrieving the SR model from the lookup table and determining whether model fine-tuning is required, and a prefetching strategy to balance the network bandwidth allocation between cloud gaming video and SR models. Experiments show that \awRe~can reduce redundant training overhead by 44\% and achieve a PSNR improvement of 1.81 dB over the baseline without additional training overhead. In addition, we implement our method on mobile devices to evaluate end-to-end latency, confirming that our method meets real-time requirements. 

\clearpage
\bibliographystyle{ACM-Reference-Format}
\bibliography{reference}


\end{document}